\documentclass[journal,twoside,web]{ieeecolor}

\usepackage{tmi}

\usepackage{physics}
\usepackage{url}
\usepackage{bm}
\usepackage{cite}
\usepackage{graphicx}
\usepackage{multirow}
\usepackage[font=small]{caption}
\usepackage{mathtools}
\usepackage{amsfonts,amsmath,amssymb}
\usepackage{graphicx,url}
\usepackage{helvet}
\usepackage{tabularx}
\usepackage{color}
\usepackage{float}
\usepackage[colorlinks=true]{hyperref}
\usepackage{array}
\usepackage{textcomp}

\usepackage[normalem]{ulem}
\newcommand{\revised}[1]{%
\ifx\highlightrevisions\undefined{#1}%
\else\textcolor{red}{#1}%
\fi}
\newcommand{\revisedtwo}[1]{%
\ifx\highlightrevisionstwo\undefined{#1}%
\else\textcolor{red}{#1}%
\fi}
\newcommand{\revisedthree}[1]{%
\ifx\highlightrevisionsthree\undefined{#1}%
\else\textcolor{red}{#1}%
\fi}
\newcommand{\rnum}[1]{%
\ifx\showreviewercommentnum\undefined%
\else{[\bf{R#1}] }%
\fi}
\newcommand{\del}[1]{%
\ifx\showerased\undefined%
\else{\sout{#1}}%
\fi}

\begin{document} 
 
\title{SLfRank: Shinnar-Le-Roux Pulse Design with Reduced Energy and Accurate Phase Profiles using Rank Factorization}
\author{Frank Ong*, Zheng Zhong*, Congyu Liao, Michael Lustig, Shreyas S. Vasanawala, and John M. Pauly, \IEEEmembership{Fellow, IEEE} 

\thanks{This work was supported in part by the NIH under Grant R01EB009690. (Corresponding author: Zheng Zhong.)}                                             
\thanks{F. Ong was with the Department of Radiology, Stanford University, CA 94301 USA (e-mail: frankongh@gmail.com).} 

\thanks{Z. Zhong and C. Liao are with the Department of Radiology, Stanford University, CA 94301 USA (e-mail: zzhong21@stanford.edu and cyliao@stanford.edu).} 

\thanks{M. Lustig is with the Department of Electrical Engineering and Computer Sciences, University of California at Berkeley, Berkeley, CA 94720 USA (e-mail: mlustig@eecs.berkeley.edu).}

\thanks{S. S. Vasanawala is with the Department of Radiology, Stanford University, CA 94301 USA (e-mail: vasanawala@stanford.edu).}

\thanks{J. M. Pauly is with the Department of Electrical Engineering, Stanford University, CA 94301 USA (e-mail: pauly@stanford.edu).}

\thanks{* Frank Ong and Zheng Zhong make equal contributions to the work.}
}

\bibliographystyle{IEEEtran.bst}

\maketitle

\begin{abstract}
The Shinnar-Le-Roux (SLR) algorithm is widely used to design frequency selective pulses with large flip angles. We improve its design process to generate pulses with lower energy (by as much as 26\%) and more accurate phase profiles. 

Concretely, the SLR algorithm consists of two steps: (1) an invertible transform between frequency selective pulses and polynomial pairs that represent Cayley-Klein (CK) parameters and (2) the design of the CK polynomial pair to match the desired magnetization profiles. Because the CK polynomial pair is bi-linearly coupled, the original algorithm sequentially solves for each polynomial instead of jointly. This results in sub-optimal pulses.

Instead, we leverage a convex relaxation technique, commonly used for low rank matrix recovery, to address the bi-linearity. Our numerical experiments show that the resulting pulses are almost always globally optimal in practice. For slice excitation, the proposed algorithm results in more accurate linear phase profiles. And in general the improved pulses have lower energy than the original SLR pulses.
\end{abstract}


\section{Introduction}
\label{sec:intro}

Frequency selective radio-frequency (RF) pulses are essential components in magnetic resonance imaging (MRI). Among other functions, they are used for \revised{\rnum{1.3}\del{slice excitation, }}spectral saturation, inversion, and spin-echo refocusing. \revised{\rnum{1.3}Together with a constant slice selective gradient, they are also used for slice excitation.} Several accelerated imaging techniques, such as simultaneous multi-slice imaging, further build on advances in RF pulse design \revised{\rnum{1.5}~\cite{setsompop2012improving}}. Improvements in frequency selective pulse design can therefore benefit many applications \revised{\rnum{1.5}~\cite{liu2011radiofrequency}}.


The Shinnar-Le-Roux (SLR) algorithm~\cite{shinnar_application_1989, shinnar_synthesis_1989-1, shinnar_synthesis_1989, shinnar_use_1988, shinnar_use_1989, le_roux_exact_1988, le_roux_method_1990, le_roux_simplified_1989, pauly_parameter_1991} is widely used to design frequency selective pulses with large flip angles. It vastly simplifies the highly non-linear pulse design problem by mapping RF pulses to pairs of polynomial that represent Cayley-Klein (CK) parameters. Pulse designers can then solve for polynomial pairs using filter design algorithms and convert them back. Users can also incorporate pulse energy as a design objective, which is crucial when the specific absorption rate (SAR) is a limiting factor.

However, there is a subtle issue in the current SLR design process: the algorithm does not jointly design the CK polynomial pair. This can lead to sub-optimal pulses with higher energy and inaccurate phase profiles.

The main challenge preventing joint recovery is that CK parameters are bi-linearly coupled. For example, the transverse magnetization $M_{\mathrm{xy}}$ is related to the CK parameters, $\alpha$ and $\beta$, as $M_{\mathrm{xy}} = 2 \alpha^* \beta$. To bypass the bi-linear coupling, the SLR algorithm first converts constraints on $M_{\mathrm{xy}}$ to constraints on $\beta$. It then finds a $\beta$ to satisfy the constraints and solves for an $\alpha$ to minimize pulse energy. The conversion between constraints on magnetization profiles and on $\beta$ is not exact. SLR pulses can produce different phase profiles than the desired ones because the design does not account for the phase of $\alpha$. Such process also does not recover the $\beta$ that minimizes energy.

\begin{figure}
    \centering
    \includegraphics[width=\linewidth]{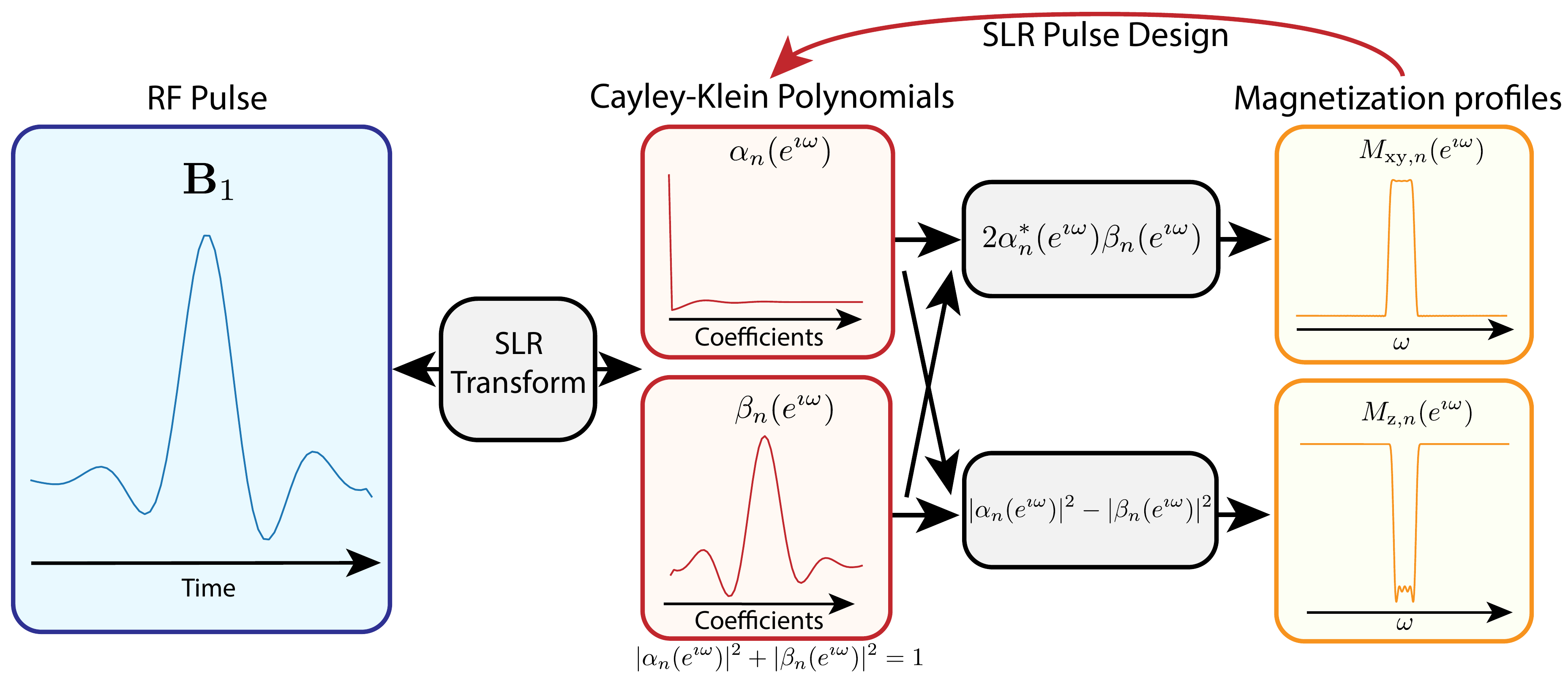}
    \caption{An overview of the SLR algorithm, which consists of two steps: (1) an invertible transform between frequency selective pulses and polynomial pairs that represent Cayley-Klein (CK) parameters and (2) the design of the CK polynomial pair to match a desired magnetization profile. Because the CK polynomial pair is bi-linearly coupled, the original algorithm sequentially solves for each polynomial instead of jointly. This results in sub-optimal pulses. Instead, we propose an improved SLR (SLfRank) design process that can jointly solve for the CK polynomial pair. The new design can specify constraints directly on magnetization profiles, and optimize both CK polynomials to minimize pulse energy. In particular, we leverage a convex relaxation technique, commonly used for low rank matrix recovery, to address the bi-linearity.}
    \label{fig:overview}
\end{figure}

To correct phase profile errors in the SLR algorithm, Barral et al.~\cite{barral_slr_2008} proposed a heuristic to alternatively solve for CK parameters. While effective, the method does not jointly optimize the CK polynomials to minimize pulse energy. Besides the SLR algorithm, other pulse design methods present different tradeoffs. A line of work using the inverse scattering transform~\cite{epstein_introduction_2004, epstein_inverse_2012, epstein_minimum_2004-1, epstein_minimum_2004, magland_discrete_2009, magland_exact_2004, magland_practical_2005} can specify constraints on magnetization profiles and minimize pulse energy. But the resulting pulses have infinite lengths. Optimal control (OC) methods~\cite{conolly_optimal_1986} have also been extensively used for pulse design. \revised{\rnum{1.1}\del{However, it is not clear in practice whether the resulting pulses achieve minimal energy. }Note that SLR technically falls into the OC framework, where we minimize an objective function subject to some constraints. The main difference between other OC methods and SLR is that they directly solve for RF magnetization or flip angles as parameters. General OC methods are more flexible in their objectives and can impose constraints on the pulse directly, such as limiting RF peak amplitude. However, they also need to solve a highly non-linear inverse problem. The SLR method, on the other hand, solves for CK polynomials, which results in a bi-linear inverse problem. And in this work, we show that we can certify the optimality of the resulting pulses.}

In particular, we propose an improved SLR design process to jointly solve for the CK polynomial pair. The new design can specify constraints directly on magnetization profiles, and optimize both CK polynomials to minimize pulse energy. We leverage a convex relaxation technique, commonly used for low rank matrix recovery~\cite{goemans_improved_1995, candes_power_2010, candes_phase_2013}, to address the bi-linearity. Although we relax the problem, the convex program allows us to check for optimality. And our numerical experiments show that the resulting pulses almost always attain the global solution in practice. Because the algorithm is based on rank factorization, we name the proposed algorithm SLfRank.

Following \cite{pauly_parameter_1991}, we use SLfRank to design pulses for excitation, inversion, saturation, and spin-echo refocusing. For slice excitation, the pulses result in more accurate linear phase profiles. And in general they have lower energy than the original SLR pulses by as much as $26\%$.

\section{Overview of the SLR algorithm}
\label{sec:slr}

Here we give an overview of the original SLR algorithm, which consists of two steps: (1) an invertible transform between frequency selective pulses and polynomial pairs that represent CK parameters and (2) the design of the CK polynomial pair to match a desired magnetization profile. 

Our work only improves the design aspect, but we describe the transform as well for completeness. We also highlight its use of quaternion representations, which provides insight into the proposed convex program.

We assume our readers are familiar with the classical Bloch equation in vector representation, but not necessarily in other forms. Therefore, we first give an introduction to the different representations used in the SLR algorithm. \revised{\rnum{1.6}Note that we assume the absence of relaxation effects throughout the paper.}

\subsection{Vectors to Quaternions to Cayley-Klein Parameters}
\label{ssec:quaternion}

\begin{figure}
    \centering
    \includegraphics[width=\linewidth]{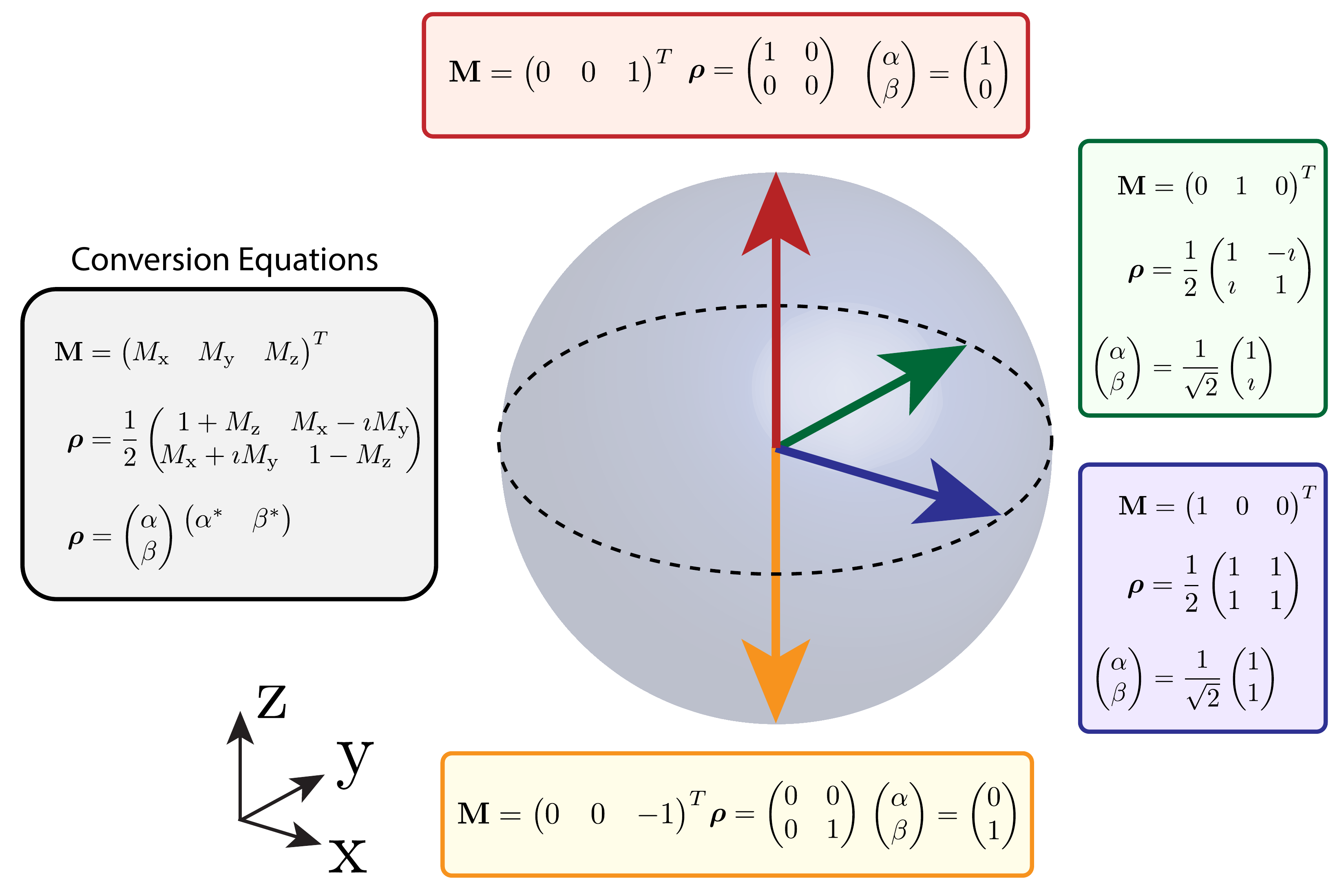}
    \caption{Illustration of different representations: magnetization vector $\mathbf{M}$, quaternion $\bm{\rho}$, and Cayley-Klein (CK) parameters \(\alpha\) and \(\beta\). We can map vectors to quaternions using an orthogonal matrix basis formed by the Pauli matrices and the identity matrix. CK parameters are simply rank-one factors of quaternions. This shows the bi-linear relationship between magnetization vectors and CK parameters.}
    \label{fig:conversion}
\end{figure}

We begin by going over the relationship between magnetization vectors and CK parameters. We use the quaternion as an intermediate representation, which allows us to map vectors to quaternions, and then to CK parameters. This path exposes the bi-linear relationship between vectors and CK parameters, which we focus on in later sections. Figure~\ref{fig:conversion} provides illustrative examples of different representations.

Concretely, let us define the following matrices:
\begin{align*}
&\mathbf{I} =   
\begin{pmatrix}
    1   &  0 \\
    0 &  1
  \end{pmatrix}
&&\bm{\sigma}_{\mathrm{x}} =   
\begin{pmatrix}
    0   &  1 \\
    1 &  0
  \end{pmatrix}
  \\
&\bm{\sigma}_{\mathrm{y}} =   
\begin{pmatrix}
    0   &  -i \\
    i &  0
  \end{pmatrix}
&&\bm{\sigma}_{\mathrm{z}} =   
\begin{pmatrix}
    1   &  0 \\
    0 &  -1
  \end{pmatrix}
\end{align*}
where \(i = \sqrt{-1}\). The matrices $\bm{\sigma}_{\mathrm{x}}$, $\bm{\sigma}_{\mathrm{y}}$, and $\bm{\sigma}_{\mathrm{z}}$ are often called the Pauli matrices. Together with the identity matrix $\mathbf{I}$, they form an orthogonal basis for 2-by-2 Hermitian matrices.

Then, for any magnetization vector $\mathbf{M} = 
\begin{pmatrix}
M_{\mathrm{x}} & M_{\mathrm{y}} & M_{\mathrm{z}}
\end{pmatrix}^T$, we can convert it to a quaternion $\bm{\rho}$ as
\begin{align}
  \bm{\rho}
  &= \frac{1}{2} (\mathbf{I} + M_{\mathrm{x}} \bm{\sigma}_{\mathrm{x}} + M_{\mathrm{y}} \bm{\sigma}_{\mathrm{y}} + M_{\mathrm{z}} \bm{\sigma}_{\mathrm{z}})
  \\
  &=
  \frac{1}{2}
  \begin{pmatrix}
    1 + M_{\mathrm{z}}  &  M_{\mathrm{x}} - i M_{\mathrm{y}} \\
    M_{\mathrm{x}} + i M_{\mathrm{y}} &  1 - M_{\mathrm{z}}
  \end{pmatrix}.
  \label{eq:m2p}
\end{align}
Note that the trace of \(\bm{\rho}\) is one by construction.

Quaternions can be seen as extensions of complex numbers, with the Pauli matrices acting like the imaginary number. And just as complex numbers can easily represent two-dimensional rotations, quaternions can compactly describe three-dimensional rotations.

In particular, if we want to rotate a quaternion $\bm{\rho}$ around an unit-norm axis \(\mathbf{u} = \begin{pmatrix} u_\mathrm{x} & u_\mathrm{y} & u_\mathrm{z} \end{pmatrix}^T\) by angle $\theta$, then defining \(\mathbf{u} \cdot \bm{\sigma} = u_{\mathrm{x}} \bm{\sigma}_{\mathrm{x}} + u_{\mathrm{y}} \bm{\sigma}_{\mathrm{y}} + u_{\mathrm{z}} \bm{\sigma}_{\mathrm{z}} \), the rotated quaternion \( \mathcal{R}_{\theta \mathbf{u}} (\bm{\rho}) \) can be expressed as
\begin{align}
  \mathcal{R}_{\theta \mathbf{u}} (\bm{\rho}) = e^{i \theta (\mathbf{u} \cdot \bm{\sigma}) / 2} \bm{\rho} e^{-i \theta (\mathbf{u} \cdot \bm{\sigma}) / 2},
  \label{eq:rotate}
\end{align}
 where the exponential is a matrix exponential given by
 \[
  e^{i \theta (\mathbf{u} \cdot \bm{\sigma}) / 2} =  \mathbf{I} \cos (\theta / 2) + i (\mathbf{u} \cdot \bm{\sigma}) \sin (\theta / 2).
 \]
 This simple rotation representation makes quaternions ideal for describing spin dynamics.

Another useful property of quaternions is that the rank of a quaternion is related to the norm of the equivalent magnetization vector. A quaternion $\bm{\rho}$ is always positive semi-definite as long as its vector representation $\mathbf{M}$ has a norm less than one. That is,
\[
  \bm{\rho} \succeq 0 \Leftrightarrow |M_{\mathrm{x}} |^2 + |M_{\mathrm{y}} |^2 + | M_{\mathrm{z}} |^2 \le 1.
\]
This can be seen from the Schur complement of $\bm{\rho}$, \( [1 - M_{\mathrm{z}} - | M_{\mathrm{xy}}|^2 / (1 + M_{\mathrm{z}}) ] / 2 \), which is non-negative if and only if $\bm{\rho}$ is positive semi-definite.

Moreover, if the vector representation has a unit norm, then the Schur complement is zero. This implies that the quaternion is rank deficient. Because the trace of the quaternion is one by construction, the rank can only be one. Therefore, we obtain
\begin{align}
  \bm{\rho} = 
  \begin{pmatrix}
      \alpha \\
      \beta
  \end{pmatrix}
  \begin{array}{@{}c@{}}
      \begin{pmatrix}
          \alpha^* & \beta^*
      \end{pmatrix} \\
    \mathstrut 
  \end{array}
  \Leftrightarrow |M_{\mathrm{x}} |^2 + |M_{\mathrm{y}} |^2 + | M_{\mathrm{z}} |^2 = 1,
  \label{eq:p2ab}
\end{align}
for some \(\alpha \in \mathbb{C}\) and \(\beta \in \mathbb{C}\) such that \(|\alpha|^2 + | \beta |^2 = 1\).

The parameters \(\alpha\) and \(\beta\) are often called the Cayley-Klein (CK) parameters. Because rotation operators in Eq.~\eqref{eq:rotate} preserve the quaternion rank, CK parameters can always represent magnetization in the absence of relaxation effects. Note also that the energy of CK parameters must sum to one. That is,
\[
\alpha \alpha^* + \beta \beta^* = 1.
\]

To convert between CK parameters and magnetization vectors, we simply match equations~\eqref{eq:m2p} and~\eqref{eq:p2ab}, and obtain
\begin{align*}
M_{\mathrm{xy}} = 2 \beta \alpha^*
~\text{and}~
M_{\mathrm{z}} = \alpha \alpha^* - \beta \beta^*,
\end{align*}
which explicitly shows the bi-linear relationship between CK parameters and magnetization vectors.

In summary, we can convert vectors to quaternions using the Pauli matrices. CK parameters are then factors of quaternions when energy is conserved.


\subsection{Forward SLR Transform}
\label{sec:forward_slr}

\begin{figure}
    \centering
    \includegraphics[width=\linewidth]{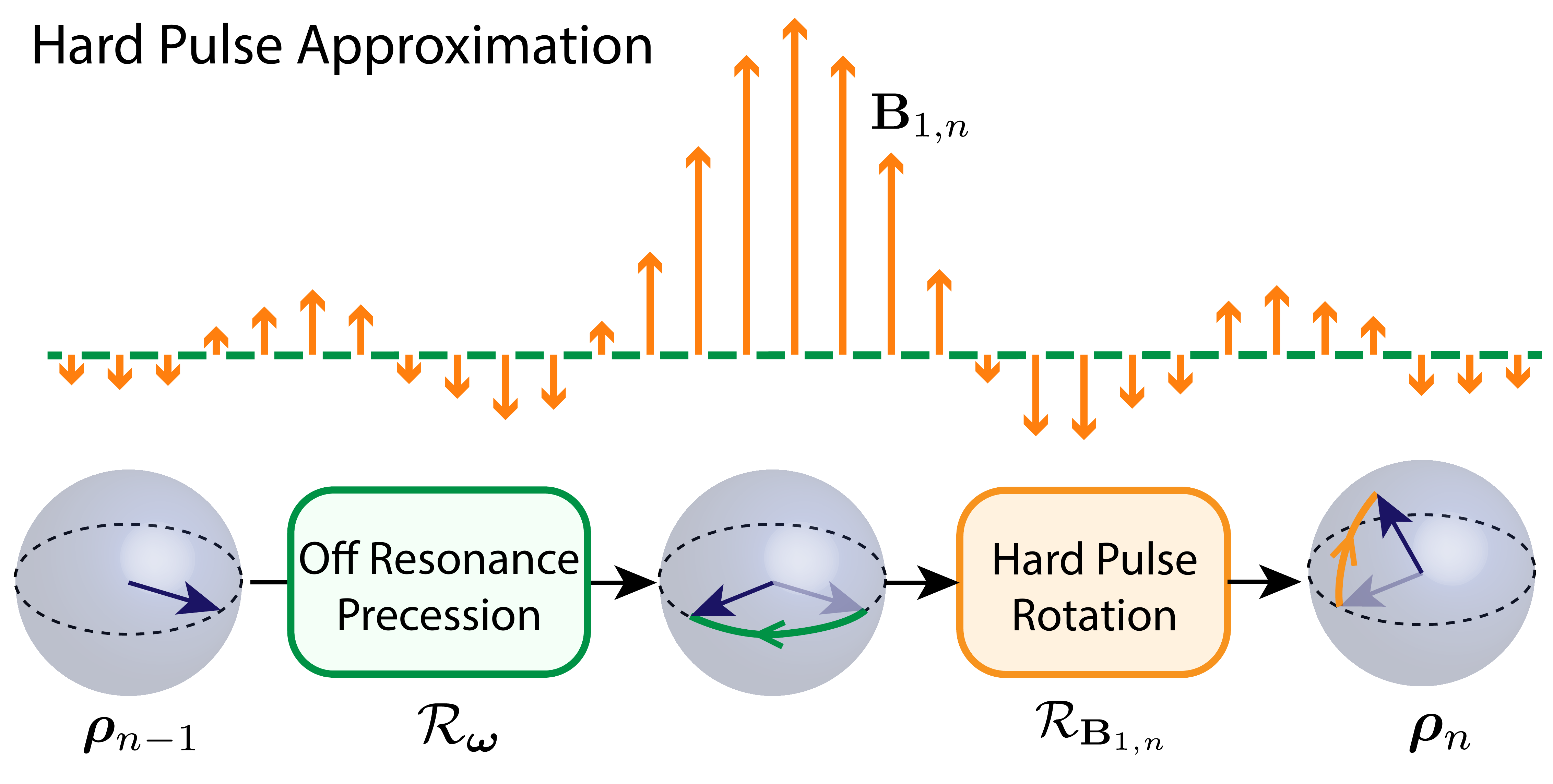}
    \caption{The SLR algorithm considers the hard pulse approximation, which discretizes the RF pulse to a sequence of impulses, also known as hard pulses. The magnetization evolution can then \revised{\rnum{4.5}be} described as a recursion of off-resonance precession followed by hard pulse rotation. Off-resonance rotates magnetization around the z-axis. And a hard pulse rotates magnetization around an axis in the transverse plane.}
    \label{fig:forward_slr}
\end{figure}

With the quaternion and CK parameter representations, we can obtain the forward SLR transform by going through a discretized version of the Bloch equation. In particular, the forward SLR transform converts any RF pulse to two polynomials that represent CK parameters. 

To do so, the forward SLR transform first considers the hard pulse approximation, which discretizes the RF pulse to a sequence of impulses, also known as hard pulses. The magnetization evolution then becomes a recursion of off-resonance precession followed by hard pulse rotation. Off-resonance rotates magnetization around the z-axis. And a hard pulse rotates magnetization around an axis in the transverse plane. Figure~\ref{fig:forward_slr} provides an illustration.

Concretely, let \(
\bm{\omega} = 
\begin{pmatrix}
0 & 0 & \omega
\end{pmatrix}^T
\) represent off-resonance with frequency $\omega$ in radian and \( \mathbf{B}_{1, n} = 
\begin{pmatrix}
B_{1, \mathrm{x}, n} & B_{1, \mathrm{y}, n} & 0
\end{pmatrix}^T\) represent the $n$th hard pulse in radian, then the quaternion after the \(n\)th hard pulse is given by
\begin{align}
    \bm{\rho}_{n} = \mathcal{R}_{\mathbf{B}_{1, n}}(\mathcal{R}_{\bm{\omega}}(\bm{\rho}_{n - 1})).
\end{align}

Using equation~\eqref{eq:rotate}, we obtain the following expressions for the rotation operators:
\[
  \mathcal{R}_{\mathbf{B}_{1, n}}(\bm{\rho}) 
  = \begin{pmatrix}
    c_n  &  -s_n^* \\
    s_n &  c_n
  \end{pmatrix} 
  \bm{\rho}
  \begin{pmatrix}
    c_n  &  s_n^* \\
    -s_n &  c_n
  \end{pmatrix}
 \]
\[
  \mathcal{R}_{\bm{\omega}}(\bm{\rho}) = 
  \begin{pmatrix}
    1  & 0 \\
    0 &  z^{-1}
  \end{pmatrix} 
  \bm{\rho}
  \begin{pmatrix}
    1  & 0 \\
    0 &  z
  \end{pmatrix}
 \]
where
\begin{align*}
    c_n &= \cos(\| \mathbf{B}_{1, n} \| / 2 ), \\ 
    s_n &= \frac{i(B_{1, \mathrm{x}, n} + i B_{1, \mathrm{y}, n})}{\| \mathbf{B}_{1, n} \|} \sin(\| \mathbf{B}_{1, n} \| / 2 ), \\
    z &= e^{i \omega}.
\end{align*}

We can further simplify the forward evolution by using CK parameters as representations. Substituting the rotation operators and using equation~\eqref{eq:p2ab}, we have
\begin{align}
    \begin{pmatrix}
        \alpha_{n} \\
        \beta_{n}
    \end{pmatrix} 
    = 
    \begin{pmatrix}
    c_n  &  -s_n^* \\
    s_n &  c_n
  \end{pmatrix} 
  \begin{pmatrix}
    1  & 0 \\
    0 &   z^{-1}
  \end{pmatrix} 
  \begin{pmatrix}
        \alpha_{n - 1} \\
        \beta_{n - 1}
    \end{pmatrix} .
    \label{eq:forward_recursion}
\end{align}

In each time step, the CK parameters gain one additional \(z\) factor. Therefore, starting from an initial CK parameter $\alpha_0 = 1$ and $\beta_0 = 0$ at equilibrium, the CK parameters after the \(n\)th hard pulse become degree-\((n - 1)\) polynomials in $z$. That is
\begin{equation}
\begin{aligned}
    \alpha_{n}(z) &= \sum_{j = 0}^{n - 1} a_{n, j}  z^{-j}, \\
    \beta_{n}(z) &= \sum_{j = 0}^{n - 1} b_{n, j}  z^{-j}.
    \label{eq:ab}
\end{aligned}
\end{equation}

Note that the energy of the CK parameters is preserved for all \(n\). That is,
\begin{align}
  \alpha_n(z) \alpha_n^*(z) + \beta_n(z) \beta_n^*(z) &= 1.
  \label{eq:trace}
\end{align}

And we can easily obtain magnetization vectors after the \(n\)th hard pulse as:
\begin{equation}
\begin{aligned}
M_{\mathrm{xy}, n}(z) &= 2 \beta_n(z) \alpha_n^*(z)
\\
M_{\mathrm{z}, n}(z) &= \alpha_n(z) \alpha_n^*(z) - \beta_n(z) \beta_n^*(z),
\label{eq:ab2m}
\end{aligned}
\end{equation}

In summary, under the hard pulse approximation, the forward SLR transform can convert any RF pulse to two polynomials that represent the CK parameters, as shown in Eq.~\eqref{eq:ab}. The polynomials must also satisfy an energy constraint in Eq.~\eqref{eq:trace}. What is remarkable is that given two polynomials with the energy constraint, we can always convert them back to a valid pulse with the inverse SLR transform.

\subsection{Inverse SLR Transform}
\label{sec:inverse_slr}

Given any polynomial pair that satisfies the energy constraint in Eq.~\eqref{eq:trace}, the inverse SLR transform can map it back to an RF pulse. It does so by considering the following backward recursion of Eq.~\eqref{eq:forward_recursion}:
\begin{align*}
    \begin{pmatrix}
        \alpha_{n - 1} \\
        \beta_{n - 1}
    \end{pmatrix} 
    &= 
  \begin{pmatrix}
    1  & 0 \\
    0 &   z
  \end{pmatrix} 
    \begin{pmatrix}
    c_n  &  s_n^* \\
    -s_n &  c_n
  \end{pmatrix} 
  \begin{pmatrix}
        \alpha_{n} \\
        \beta_{n}
    \end{pmatrix}.
\end{align*}
It then recovers \(c_n\) and \(s_n\) such that  \(\alpha_{n - 1}\) and \(\beta_{n - 1}\) represent valid CK parameters.

For \(\alpha_{n - 1}\) and \(\beta_{n - 1}\) to be valid, they need to be degree-\( (n - 2) \) polynomials and satisfy the energy constraint. Note that as long as \(| c_n |^2 + | s_n |^2 = 1\), the energy constraint is satisfied. To ensure \(\alpha_{n - 1}\) and \(\beta_{n - 1}\) are degree-\( (n - 2) \) polynomials, we need
\begin{align}
-s_n a_{n, 0} + c_n b_{n, 0} &= 0\\
c_n a_{n, n - 1} + s_n^* b_{n, n - 1} &= 0,
\end{align}
which allow us to solve for two sets of solutions for \(c_n\) and \(s_n\).

It turns out these two solutions are the same because the energy constraint in Eq.~\eqref{eq:trace} implies
\[
a_{n, n - 1} a^*_{n, 0} + b_{n, n - 1} b^*_{n, 0} = 0,
\]
where the left hand side is the leading coefficient of the polynomial \(\alpha_n(z) \alpha_n^*(z) + \beta_n(z) \beta_n^*(z)\).

Therefore, we can recover the parameters as:
\begin{align*}
c_n &= \frac{a_{n, 0}}{\sqrt{|a_{n, 0}|^2 + |b_{n, 0}|^2}} 
= \frac{b_{n, n - 1}}{\sqrt{|a_{n, n - 1}|^2 + |b_{n, n - 1}|^2}} \\
s_n &= \frac{b_{n, 0}}{\sqrt{|a_{n, 0}|^2 + |b_{n, 0}|^2}}
= \frac{-a_{n, n - 1}^*}{\sqrt{|a_{n, n - 1}|^2 + |b_{n, n - 1}|^2}}
\end{align*}

And we can obtain the \(n\)th hard pulse as,
\begin{align*}
\phi_n &= 2 \arctan(|s_n / c_n|),\\
\theta_n &= \angle( -\imath s_n c_n^*), \\
B_{1, n, \mathrm{x}} &= \phi_n \cos(\theta_n), \\
B_{1, n, \mathrm{y}} &= \phi_n \sin(\theta_n).
\end{align*}

\subsection{Original SLR Design Process}
\label{sec:original_design}

With the SLR transform, the highly non-linear pulse design problem becomes equivalent to designing the CK polynomial pair. The remaining challenge is the bi-linear coupling between \(\alpha_n\) and \(\beta_n\). The original SLR algorithm bypasses this by sequentially solving for each variable. It first converts design constraints on magnetization profiles to constraints on \(\beta_n\). Then, it recovers an \(\alpha_n\) that minimizes the pulse energy. In the following, we highlight some advantages and disadvantages of this approach.

There are certain constraints that can be directly expressed in \(\beta_n\). \cite{pauly_parameter_1991} shows that for spin-echo refocusing pulses, the effective transverse magnetization after crusher gradients can be expressed as
\begin{align*}
    M_{\mathrm{xy}, n}(z) &= \beta_n^2(z).
\end{align*}
For single band designs, any constraints on \(M_{\mathrm{xy}, n}\) can then be converted to constraints on \(\beta_n\) up to a global sign change, which does not affect the resulting pulse.

However, for other pulses, we cannot directly translate constraints. Instead, using Eqs.~\eqref{eq:trace} and~\eqref{eq:ab2m}, we only have the following relationships:
\begin{align*}
    |M_{\mathrm{xy}, n}(z)|^2 &= 4 | \beta_n(z) |^2 (1 - |\beta_n(z) |^2),\\
    M_{\mathrm{z}, n}(z) &= 1 - 2 |\beta_n(z)|^2.
\end{align*}
\cite{pauly_parameter_1991} shows that we can use these equations to specify ripple parameters in min-max filter designs. However, note that we cannot express transverse magnetization phase in terms of \(\beta_n\).

When the user does not require a particular transverse magnetization phase profile, the original SLR algorithm has more flexibility. One choice is a minimum phase polynomial for \( \beta_n \). This is equivalent to selecting a \(\beta_n\) polynomial that maximizes its first coefficient \(\Re{b_{n, 0}}\). Using linear approximation, we can interpret this as maximizing the last hard pulse:
\[
\Re{b_{n, 0}} = c_1 \ldots c_{n - 1} s_n \approx B_{1, y, n} / 2.
\]

However, when the user wants a specific transverse magnetization phase response, such as in slice excitation, the original SLR algorithm cannot find a corresponding \(\beta_n\) polynomial to do so. Instead, it relies on an approximation that \(\alpha_n\) does not contribute much phase and finds a \(\beta_n\) to account for all of the phase of \(M_{\mathrm{xy}}\). But \(\alpha_n\) always contributes some phase. Therefore, the resulting profile in general deviates from the desired one.

Once we obtain a \(\beta_n\), the original design process recovers the \(\alpha_n\) polynomial by minimizing pulse energy. In particular, the first polynomial coefficient of \(\alpha_n\) acts as a proxy for pulse energy. Using \(c_j \approx 1 - \| \mathbf{B}_{1, j} \|^2 / 8\), we have
\[ 
a_{n, 0} = c_1 c_2 \ldots c_n \approx 1 - \frac{1}{8} \sum_{j = 1}^n \| \mathbf{B}_{1, j} \|^2.
\]

With fine enough discretization, maximizing \(a_{n, 0}\) minimizes pulse energy. We can then find the corresponding \(\alpha_n\) by solving for a minimum phase filter.

On the other hand, the \(\beta_n\) polynomial design does not take pulse energy into consideration. It is possible a different \(\beta_n\) that satisfies the constraints results in lower pulse energy. Indeed, our improved SLR design shows that jointly designing \(\alpha_n\) and \(\beta_n\) reduces pulse energy in general.

\section{SLfRank: Improved SLR Pulse Design with Rank Factorization}
\label{sec:design}

We propose an improved SLR design, named SLfRank, to jointly recover the CK polynomials with a convex program. The new design can specify constraints directly on magnetization profiles, and optimize both CK polynomials to minimize pulse energy. To derive the optimization problem, we first show that all constraints on the CK polynomials can be represented as linear equations on a rank-one matrix. Then we relax matrix rank constraints to positive semi-definite matrix constraints to obtain a convex program.

Concretely let us define the following vectors:
\begin{align*}
\bm{\psi}(z) &= 
\begin{pmatrix}
1 &
 z &
\ldots &
 z^{n - 1}
\end{pmatrix}^T  \in \mathbb{C}^{n}\\
\mathbf{a} &= 
\begin{pmatrix}
a_{n, 0} &
 a_{n, 1} &
\ldots &
 a_{n, n - 1}
\end{pmatrix}^T  \in \mathbb{C}^{n} \\
\mathbf{b} &= 
\begin{pmatrix}
b_{n, 0} &
 b_{n, 1} &
\ldots &
 b_{n, n - 1}
\end{pmatrix}^T \in \mathbb{C}^{n}
\end{align*}
\revised{\rnum{2.2}where $^T$ denotes the transpose operation without complex conjugation. The vectors $\mathbf{a}$ and $\mathbf{b}$ represent the polynomial coefficients for $\alpha_n$ and $\beta_n$ respectively. $\bm{\psi}(z)$ represents the complex exponentials.}

Then the CK polynomials can be expressed as,
\begin{align*}
\alpha_n(z) &= \bm{\psi}^*(z) \mathbf{a}, \\
\beta_n(z) &= \bm{\psi}^*(z) \mathbf{b},
\end{align*}
\revised{\rnum{2.2}where $^*$ denotes the Hermitian transpose operation.}



Let us further define $\mathbf{P} \in \mathbb{C}^{2n \times 2n}$ as the outer product of the CK polynomial coefficients \revised{\rnum{3.3, 4.1} and partition the matrix into submatrices $\mathbf{P}_{aa}, \mathbf{P}_{ba}, \mathbf{P}_{ab}, \mathbf{P}_{bb} \in \mathbb{C}^{n \times n}$ as follows}:
\[
    \mathbf{P}
     =
      \begin{pmatrix}
        \mathbf{P}_{aa} & \mathbf{P}_{ab} \\
        \mathbf{P}_{ba} & \mathbf{P}_{bb} \\
      \end{pmatrix}
    = 
      \begin{pmatrix}
          \mathbf{a} \\
          \mathbf{b}
      \end{pmatrix}
      \begin{array}{@{}c@{}}
          \begin{pmatrix}
              \mathbf{a}^* & \mathbf{b}^*
          \end{pmatrix} \\
        \mathstrut 
      \end{array}
\]

Then, we can express the energy constraint in Eq.~\eqref{eq:trace} and magnetization profiles Eq.~\eqref{eq:ab2m} in terms of $\mathbf{P}$ as:
\begin{align*}
   1 &= \bm{\psi}^*(z)  (\mathbf{P}_{aa} + \mathbf{P}_{bb}) \bm{\psi}(z) \\
    M_{\mathrm{xy}, n}(z) &= 2 \bm{\psi}^*(z)  \mathbf{P}_{ba} \bm{\psi}(z) \\
    M_{\mathrm{z}, n}(z) &= \bm{\psi}^*(z)  (\mathbf{P}_{aa} - \mathbf{P}_{bb}) \bm{\psi}(z).
\end{align*}

The above equations show that we can express all constraints on the CK polynomials as linear equations on a rank-one matrix $\mathbf{P}$. We can also easily change them to impose inequality constraints on magnetization profiles.

Taking a step back, the matrix \(\mathbf{P}\) we have just formed essentially represent the underlying quaternion \(\rho_n(z)\). In particular, we have
\[
    \rho_n(z)
     =
          \begin{pmatrix}
              \bm{\psi}^*(z) & \mathbf{0} \\
              \mathbf{0} & \bm{\psi}^*(z)\\
          \end{pmatrix}
          \mathbf{P}
          \begin{pmatrix}
              \bm{\psi}(z) & \mathbf{0} \\
              \mathbf{0} & \bm{\psi}(z)\\
          \end{pmatrix}
\]
Therefore, one way to interpret the proposed design is that we solve for quaternions instead of CK parameters. And the two representations become equivalent when the quaternion is rank-one.

Optimizing over rank-one matrices is in general non-convex. A common strategy is to relax the rank constraint into a positive semi-definite matrix constraint. Concretely, we can relax the constraint as follows,
\[
    \mathbf{P}
    \succeq
      \begin{pmatrix}
          \mathbf{a} \\
          \mathbf{b}
      \end{pmatrix}
      \begin{array}{@{}c@{}}
          \begin{pmatrix}
              \mathbf{a}^* & \mathbf{b}^*
          \end{pmatrix} \\
        \mathstrut 
      \end{array},
\]
\revised{\rnum{3.2, 3.3}which is convex. This can be seen using the properties of the Schur complement. In particular, the constraint is equivalent to:
\[
\mathbf{X} = 
      \begin{pmatrix}
            1 & \mathbf{a}^* & \mathbf{b}^* \\
          \mathbf{a} & \mathbf{P}_{aa} & \mathbf{P}_{ab} \\
          \mathbf{b} & \mathbf{P}_{ba} & \mathbf{P}_{bb} \\
      \end{pmatrix}
      \succeq 0.
\]
Another way of looking at the convex relaxation is that we relax the constraint that $\mathbf{X}$ is a rank-1 matrix to that $\mathbf{X}$ being a positive semi-definite matrix. Such relaxation has been applied in many other applications, such as max-cut~\cite{goemans_improved_1995}, matrix completion~\cite{candes_power_2010}, and phase retrieval~\cite{candes_phase_2013}.
}



Similar to the original SLR algorithm, we maximize \(a_{n, 0}\) to minimize pulse energy and \(b_{n, 0}\) to generate minimum phase pulses. Putting everything together, we obtain the following optimization problem to design the CK polynomials:
\begin{equation*}
\begin{aligned}
&\underset{\revised{\rnum{3.3}\mathbf{a}, \mathbf{b}, \mathbf{P}}}{\text{max}}
&&\Re(a_{n, 0}) + \lambda_{\text{mp}} \Re(b_{n, 0})\\
&\text{s.t.}
&&\revised{\begin{pmatrix}
            1 & \mathbf{a}^* & \mathbf{b}^* \\
          \mathbf{a} & \mathbf{P}_{aa} & \mathbf{P}_{ab} \\
          \mathbf{b} & \mathbf{P}_{ba} & \mathbf{P}_{bb} \\
      \end{pmatrix}}
    \succeq 0, \\
&&&\bm{\psi}^*(e^{i \omega})  (\mathbf{P}_{aa} + \mathbf{P}_{bb}) \bm{\psi}(e^{i \omega}) = 1,\\
&&& | 2 \bm{\psi}^*(e^{i \omega})  \mathbf{P}_{ba} \bm{\psi}(e^{i \omega}) - M_{\mathrm{xy}, n}(e^{i \omega}) | \leq \delta_{\mathrm{xy}}(e^{i \omega}),\\
&&& | \bm{\psi}^*(e^{i \omega})  (\mathbf{P}_{aa} - \mathbf{P}_{bb}) \bm{\psi}(e^{i \omega}) - M_{\mathrm{z}, n}(e^{i \omega}) | \leq \delta_{\mathrm{z}}(e^{i \omega}),\\
&&&  \revised{\rnum{1.8}\text{(The following constraint is for spin-echo refocusing)}} \\
&&& | \bm{\psi}^*(e^{i \omega})  \mathbf{b} - \beta_n(e^{i \omega}) | \leq \delta_{\beta}(e^{i \omega}) ,
\end{aligned}
\end{equation*}
where \(\omega\) goes from \(-\pi\) to \(\pi\), \(\lambda_{\text{mp}}\) enforces minimum phase conditions, \(\delta_{\mathrm{xy}}\), \(\delta_{\mathrm{z}}\), and \(\delta_{\beta}\) are user-defined error parameters, and \(M_{\mathrm{xy}, n}\) and \(M_{\mathrm{z}, n}\) represent the desired magnetization profiles. \revised{\rnum{3.3}Note that $a_{n, 0}$ and $b_{n, 0}$ are the first elements of $\mathbf{a}$ and $\mathbf{b}$ respectively and $\mathbf{P}_{aa}, \mathbf{P}_{ba}, \mathbf{P}_{ab}, \mathbf{P}_{bb}$ are submatrices of $\mathbf{P}$.} \revised{\rnum{1.8}Also the constraints on $\mathbf{b}$ are only imposed for spin-echo refocusing.}

\revised{\rnum{3.3}Table~\ref{tab:params1} and~\ref{tab:params2} contain the \(\delta\) parameters used for different pulse types. For bands that are not specified, the \(\delta\) parameters are set to 1 and the desired profiles are set to 0. } 
\begin{table*}[]
\centering
\caption{Parameters for excitation, inversion, and saturation pulse designs.}
\begin{tabular}{c|cccc|cccc}
                       & \multicolumn{4}{c|}{Pass-band with ripple $\delta_1$}                                                                     & \multicolumn{4}{c}{Stop-band with ripple $\delta_2$}                                               \\
Pulse (phase)     & $M_{\mathrm{xy}}$                      & $\delta_{\mathrm{xy}}$        & $M_{\mathrm{z}}$ & $\delta_{\mathrm{z}}$         & $M_{\mathrm{xy}}$ & $\delta_{\mathrm{xy}}$        & $M_{\mathrm{z}}$ & $\delta_{\mathrm{z}}$       \\ \hline
Excitation (linear)    & $ e^{-i \omega (n + 1)/ 2}$ & $\delta_1$                    & 0                & $\sqrt{1 - (1 - \delta_1)^2}$ & 0                 & $\delta_2$                    & 1                & $1 - \sqrt{1 - \delta_2^2}$ \\
Excitation (min.) & 0                                      & 1                             & 0                & $\sqrt{1 - (1 - \delta_1)^2}$ & 0                 & $\delta_2$                    & 1                & $1 - \sqrt{1 - \delta_2^2}$ \\
Inversion (min.)  & 0                                      & $\sqrt{1 - (1 - \delta_1)^2}$ & -1               & $\delta_1$                    & 0                 & $\sqrt{1 - (1 - \delta_2)^2}$ & 1                & $\delta_2$                  \\
Saturation (max.) & 0                                      & 1                             & 0                & $\delta_1$                    & 0                 & 1                             & 1                & $\delta_2$                 
\end{tabular}

\label{tab:params1}
\end{table*}

Although we relax the problem, the convex program allows us to check for optimality with the solution rank. If the resulting matrix \(\mathbf{P}\) is close to rank-one, then it is close to being globally optimal. And if it is exactly rank-one, then we have recovered a global minimum. In the next section, our numerical experiments will show that convex program almost always attains the global solution in practice.

Finally, the convex program in its current form imposes infinitely many constraints due to the continuous nature of \(\omega\). There are two ways to address this. One way is to finely sample frequencies and only impose constraints on the finite set. We opt for this strategy for inequalities because it is simpler. And slight violation outside the finite set is often tolerable in practice. Another way is to convert the constraints into linear equations on positive semi-definite matrices as shown in~\cite{dumitrescu_positive_2007}. In particular, the constraint \(\bm{\psi}^*(e^{i \omega})  (\mathbf{P}_{aa} + \mathbf{P}_{bb}) \bm{\psi}(e^{i \omega}) = 1\) for all \(\omega\) is equivalent to
\begin{equation}
\sum_{i, j : i - j = k} (\mathbf{P}_{aa} + \mathbf{P}_{bb})_{ij} = \begin{cases}
    1,& \text{if } k = 0\\
    0,              & \text{otherwise}.
\end{cases}
\label{eq:enery_con}
\end{equation}
for \(k = -n, -n + 1, \ldots, n\). \revised{\rnum{4.8}We have included a proof sketch in the Supplementary Materials for completeness.} We opt for this conversion for this constraint because it ensures the resulting solution exactly satisfies the energy constraint.

\revised{\rnum{2.4, 3.3}Once we obtain CK polynomial coefficients $\mathbf{a}$ and $\mathbf{b}$, we can then use the inverse SLR transform in Section~\ref{sec:inverse_slr} to recover the hard pulses $\mathbf{B}_{1, 1}, \ldots, \mathbf{B}_{1, n}$.}

\section{Numerical Experiments}

We use the proposed algorithm to design pulses for excitation, inversion, saturation, and spin-echo refocusing. All examples shown here have time-bandwidths (TBW) of 8, maximum absolute errors of 1\%, and \(n=64\). \revised{\rnum{3.4}We follow \cite{pauly_parameter_1991} to choose the same ripple, passband, stopband and transition width parameters for both SLR and SLfRank.} The inequality constraints are imposed on 960 uniformly sampled points in the frequency domain. For minimum phase pulses, \(\lambda_{\text{mp}}\) is set to one. Maximum phase pulses are generated as time-reversed minimum phase pulses. For all other pulses, \(\lambda_{\text{mp}}\) is set to zero. \revised{\rnum{1.2}We have also generated more comparisons in the Supplementary Materials with TBW=4 and 10, along with $n=100$.}

\begin{table*}[]
\centering
\caption{Parameters for spin-echo refocusing pulse designs.}
\begin{tabular}{c|cc|cc}
                   & \multicolumn{2}{c|}{Pass-band with ripple $\delta_1$}              & \multicolumn{2}{c}{Stop-band with ripple $\delta_2$} \\
Pulse (phase) & $\beta$                          & $\delta_{\beta}$                & $\beta$              & $\delta_{\beta}$              \\ \hline
Spin-Echo (zero)   & $- e^{-i \omega (n + 1) / 2}$ & $(1 - \sqrt{1 - \delta_1}) / 2$ & 0                    & $\sqrt{\delta_2}$            
\end{tabular}

\label{tab:params2}
\end{table*}

Because the optimization problem is convex, any solver can reach a global minimum. We first verify the program with CVXPY~\cite{diamond_cvxpy_2016}, which uses interior-point solvers, with \(n=16\). While CVXPY is accurate, it can be quite slow for large \(n\). For the final pulses displayed in this manuscript, we use the primal dual hybrid gradient algorithm~\cite{chambolle_first-order_2011} in SigPy~\cite{ong_sigpy:_2019} with \(n=64\). We use 20000 iterations to ensure the algorithm converges. We compare SLfRank pulses to the original SLR pulses generated with SigPy.RF~\cite{martin_sigpyrf_2020}. We compute the pulse energy, defined as the sum of squares of the pulses, and peak amplitude for comparison.

In the spirit of reproducible research, we provide a software package to reproduce the results described in this paper. The software package can be downloaded from:

\begin{center}
\url{https://github.com/MRSRL/slfrank}
\end{center}

\begin{figure}
    \centering
    \includegraphics[width=\linewidth]{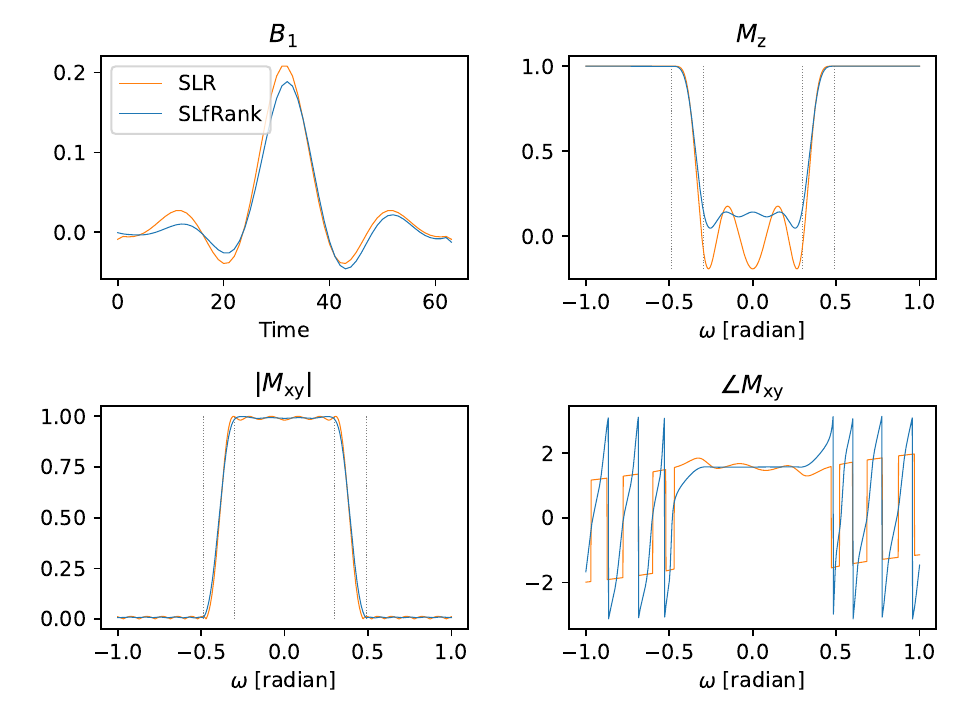}
    \caption{Linear phase excitation pulses and their magnetization profiles after \revised{\rnum{2.5}gradient} refocusing\revised{\rnum{2.6, 3.4}, with band boundaries denoted by dotted gray lines}. The SLfRank pulse has a much flatter phase response than the original one after \revised{\rnum{2.5}gradient} refocusing. Pulse energy is reduced from 0.318 to 0.259 (18.6\%) and peak is reduced from 0.208 to 0.189 (9.1\%). Note that the SLfRank pulse is asymmetric, whereas the SLR pulse is symmetric. This shows that the proposed design compensates for the phase of \(\alpha\) to generate a linear phase profile.}
    \label{fig:excitation}
\end{figure}

\begin{figure}
    \centering
    \includegraphics[width=\linewidth]{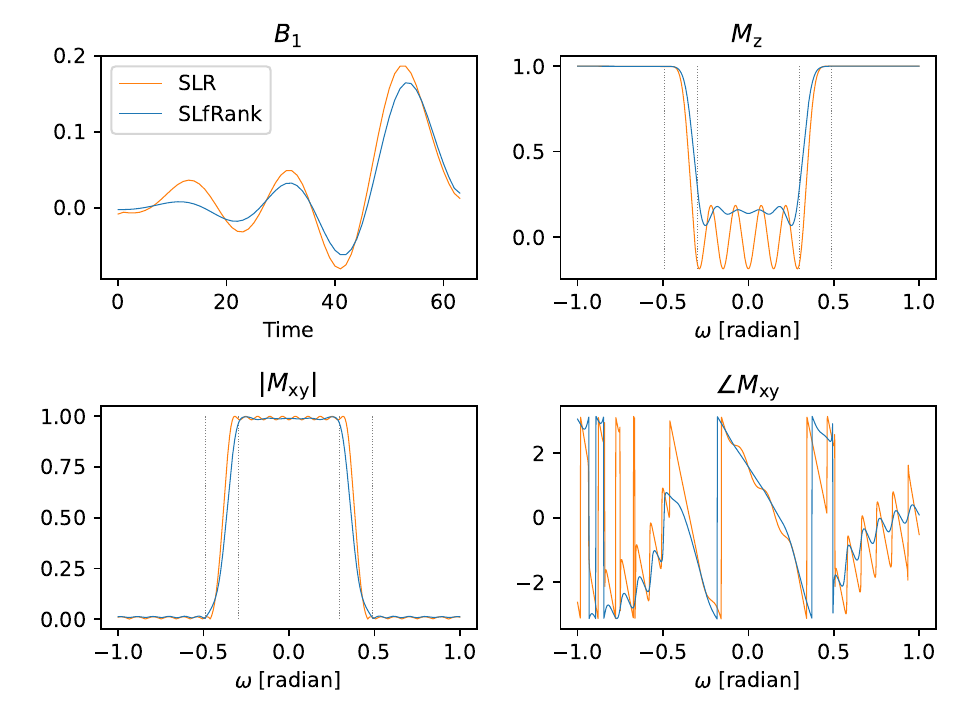}
    \caption{Minimum phase slice selection pulses and their magnetization profiles\revised{\rnum{2.6, 3.4}, with band boundaries denoted by dotted gray lines}. Pulse energy is reduced from 0.318 to 0.234 (26.4\%) and peak is reduced from 0.187 to 0.165 (11.8\%).}
    \label{fig:excitation-minphase}
\end{figure}

\begin{figure}
    \centering
    \includegraphics[width=\linewidth]{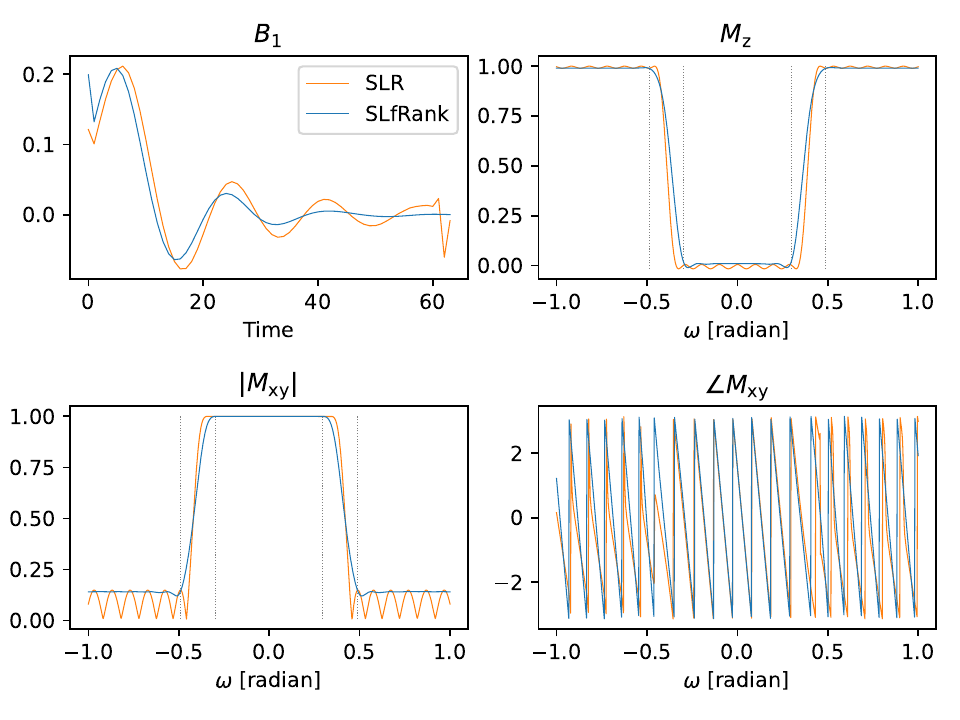}
    \caption{Maximum phase saturation pulses and their magnetization profiles\revised{\rnum{2.6, 3.4}, with band boundaries denoted by dotted gray lines}. Pulse energy is reduced from 0.352 to 0.333 (5.40\%) and peak is reduced from 0.212 to 0.208 (1.89\%). The SLfRank pulse also has fewer discontinuities at the end of the pulse, commonly known as Connolly wings, when compared to the SLR pulse.}
    \label{fig:saturation}
\end{figure}

\begin{figure}
    \centering
    \includegraphics[width=\linewidth]{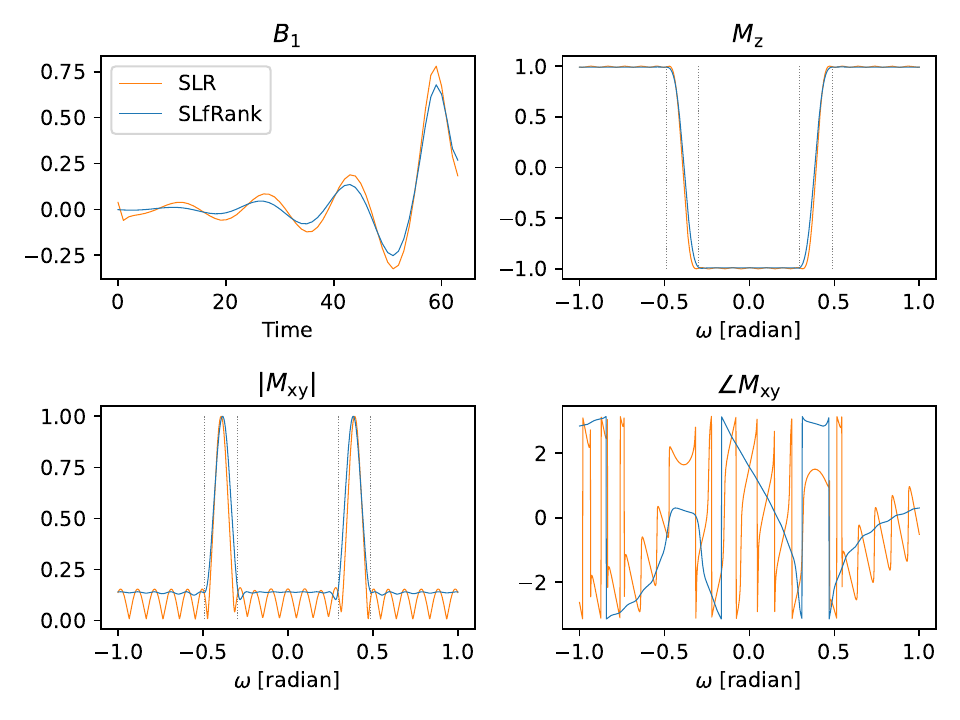}
    \caption{Minimum phase inversion pulses and their magnetization profiles\revised{\rnum{2.6, 3.4}, with band boundaries denoted by dotted gray lines}. Pulse energy is reduced from 3.00 to 2.31 (23.0\%) and peak is reduced from 0.781 to 0.679 (13.1\%).}
    \label{fig:inversion}
\end{figure}

\begin{figure}
    \centering
    \includegraphics[width=\linewidth]{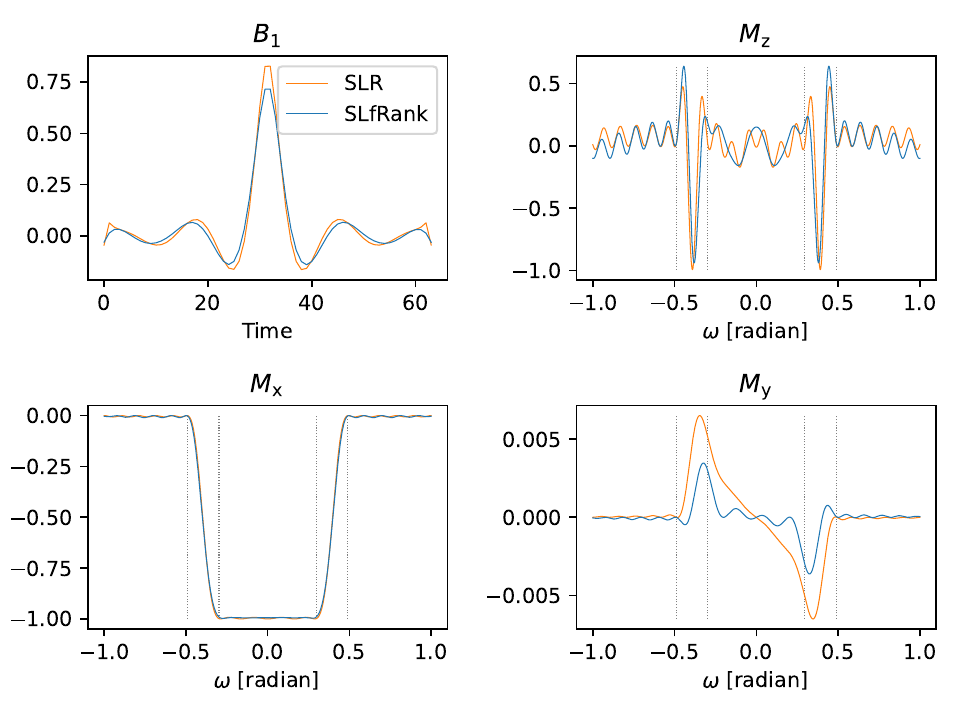}
    \caption{Spin-echo refocusing pulses and their magnetization profiles\revised{\rnum{2.6, 3.4}, with band boundaries denoted by dotted gray lines}. Pulse energy is reduced from 2.74 to 2.23 (18.6\%) and peak is reduced from 0.827 to 0.716 (13.4\%). The SLfRank pulse also has fewer discontinuities near the edges when compared to the SLR pulse.}
    \label{fig:spin-echo}
\end{figure}

Figures~\ref{fig:excitation},~\ref{fig:excitation-minphase},~\ref{fig:saturation},~\ref{fig:inversion}, and~\ref{fig:spin-echo} show the resulting pulses and their magnetization profiles. In all cases except for the minimum phase excitation pulse, the convex relaxation is tight, that is the resulting solution \(\mathbf{P}\) has rank one and
\[
\mathbf{P} 
    =
      \begin{pmatrix}
          \mathbf{a} \\
          \mathbf{b}
      \end{pmatrix}
      \begin{array}{@{}c@{}}
          \begin{pmatrix}
              \mathbf{a}^* & \mathbf{b}^*
          \end{pmatrix} \\
        \mathstrut 
      \end{array}.
\]
For the minimum phase excitation pulse, the relaxation is still quite accurate with a 0.004 \(\ell 2\) norm difference between \(\mathbf{P}\) and the outer product of \(\mathbf{a}\) and \(\mathbf{b}\).

Figure~\ref{fig:excitation} shows the linear phase excitation pulses and their magnetization profiles after refocusing. The SLfRank pulse has a much flatter phase response after refocusing than the SLR pulse. Pulse energy is reduced from 0.318 to 0.259 (18.6\%) and peak is reduced from 0.208 to 0.189 (9.1\%). Note that the SLfRank pulse is asymmetric, whereas the SLR pulse is symmetric. This shows that the proposed design compensates for the phase of \(\alpha\) to generate a linear phase profile.

Figure~\ref{fig:excitation-minphase} shows the minimum phase excitation pulses and their magnetization profiles. Pulse energy is reduced from 0.318 to 0.234 (26.4\%) and peak is reduced from 0.187 to 0.165 (11.8\%).

Figure~\ref{fig:saturation} shows the maximum phase saturation pulses and their magnetization profiles. Pulse energy is reduced from 0.352 to 0.333 (5.40\%) and peak is reduced from 0.212 to 0.208 (1.89\%). The SLfRank pulse also has fewer discontinuities at the end of the pulse, commonly known as Connolly wings, when compared to the SLR pulse.

Figure~\ref{fig:inversion} shows the minimum phase inversion pulses and their magnetization profiles.  Pulse energy is reduced from 3.00 to 2.31 (23.0\%) and peak is reduced from 0.781 to 0.679 (13.1\%).

Finally, figure~\ref{fig:spin-echo} shows the spin-echo refocusing pulses and their magnetization profiles. Pulse energy is reduced from 2.74 to 2.23 (18.6\%) and peak is reduced from 0.827 to 0.716 (13.4\%). The SLfRank pulse also has fewer discontinuities near the edges when compared to the SLR pulse.

\section{Phantom Experiment}

\begin{figure}
    \centering
    \includegraphics[width=\linewidth]{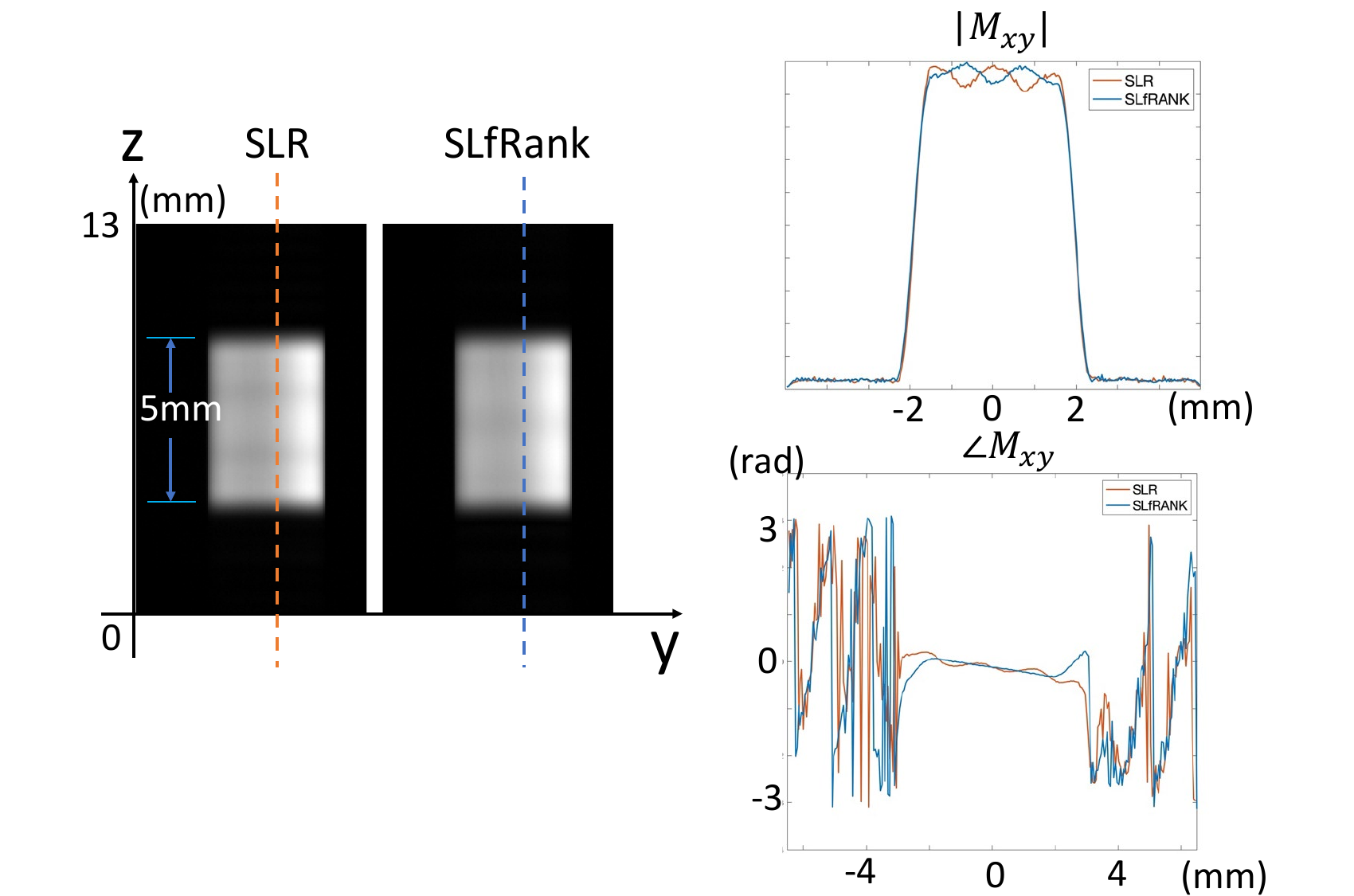}
    \caption{\revised{\rnum{3.1}Excitation slice profile using the linear phase excitation pulse generated from SLR and SLfRank, respectively. The measured slice profile matches well with the numerical simulation result. The phase response of SLfRank pulse is more linear than SLR.}}
    \label{fig:sl-profile}
\end{figure}

\revised{\rnum{3.1}To demonstrate the technical feasibility of the proposed algorithm, we performed a phantom experiment to measure the slice profile of a linear phase excitation pulse designed using both SLR and SLfRank. We also compared it with numerical results. The experiment was performed on a GE 3T scanner with a 32-channel head coil, using a custom-built GRE sequence by changing the frequency-encoding gradient to slice-selection direction. As shown in Figure~\ref{fig:sl-profile}, the measured slice profiles matched well with the numerical experiment. Note that the phase profile of SLfRank pulse is flatter than that of SLR pulse. }

\begin{figure}
    \centering
    \includegraphics[width=\linewidth]{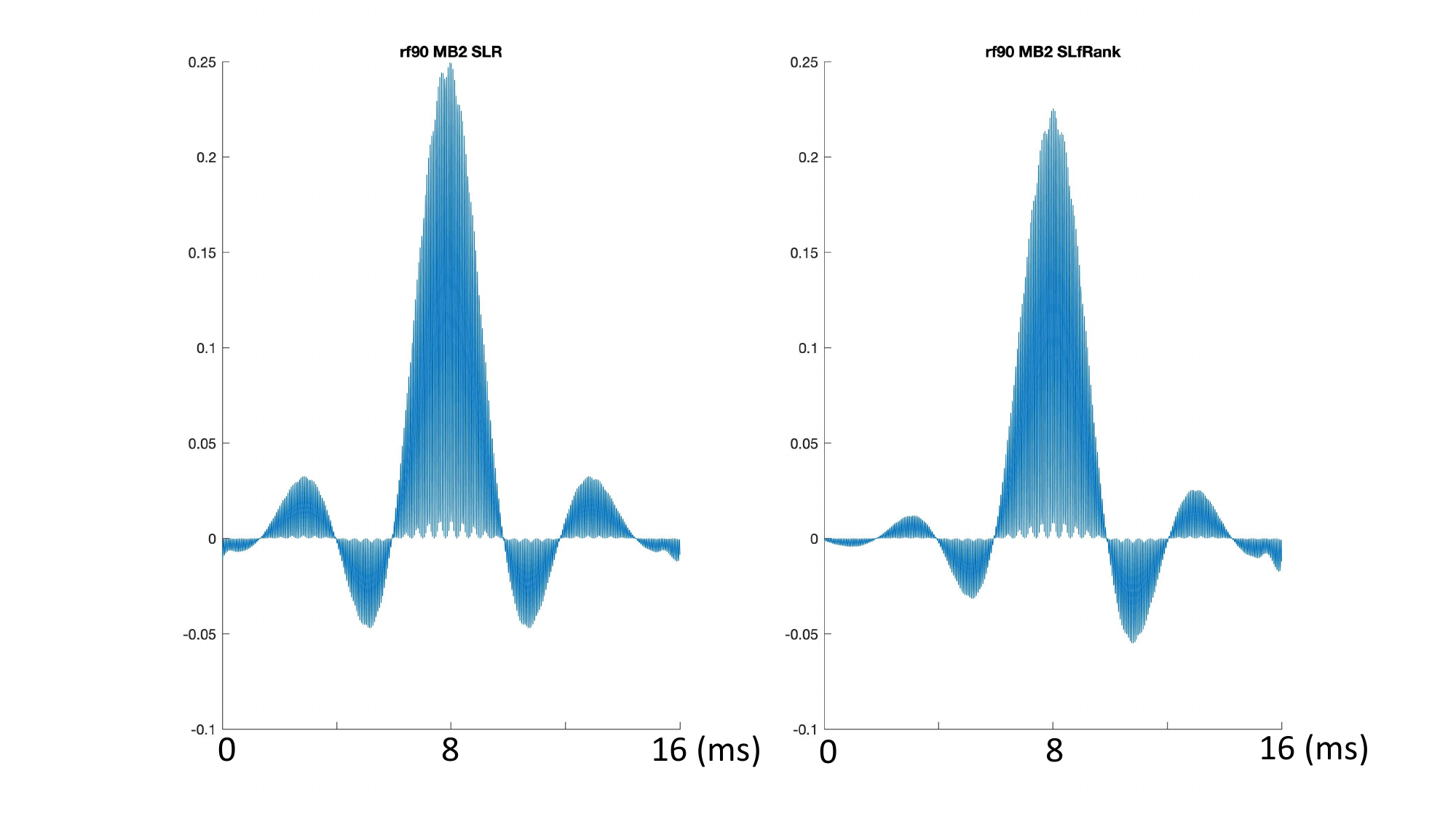}
    \caption{\revised{\rnum{3.1}The multi-band excitation pulses (MB factor = 2) designed using SLR and SLfRank, respectively. The energy of the two pulses is 0.142 and 0.115, respectively, indicating a 19\% reduction of the RF energy.}}
    \label{fig:sms-rf-mb2}
\end{figure}

\begin{figure}
    \centering
    \includegraphics[width=\linewidth]{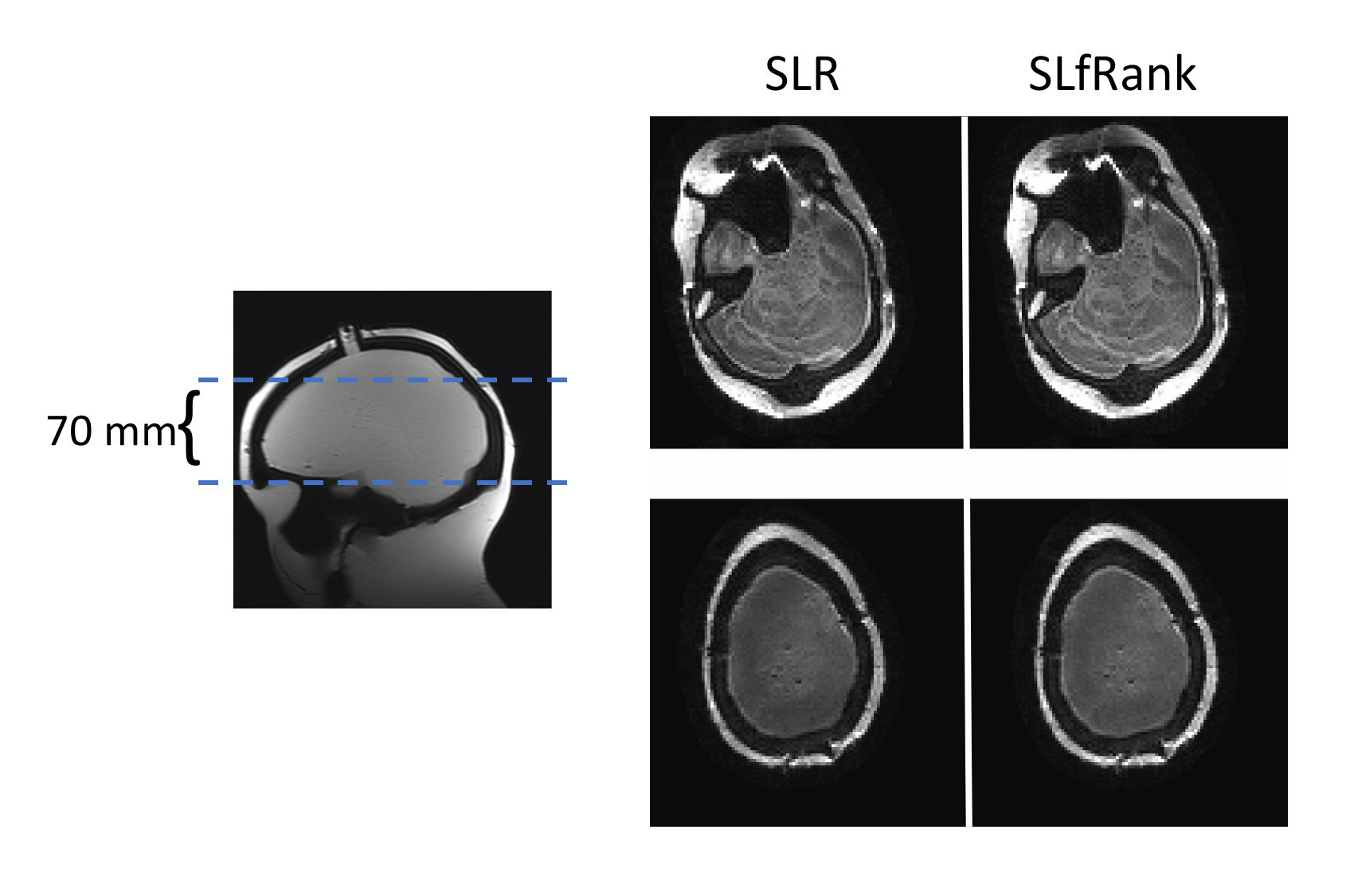}
    \caption{\revised{\rnum{3.1}Brain water phantom images acquired using the MB pulses in Figure~\ref{fig:sms-rf-mb2}. The two slices were simultaneously excited with 70 cm apart.}}
    \label{fig:mb2-phantom}
\end{figure}

\revised{\rnum{2.8, R3.1}To further demonstrate its application, we implemented multi-band RF pulse (Figure ~\ref{fig:sms-rf-mb2}) and tested it on a brain water phantom (Figure ~\ref{fig:mb2-phantom}). Specifically, the RF pulse was designed with a multi-band factor of 2, slice thickness of 2 mm and 70 mm apart. The RF pulse length was 16 ms, which was then incorporated into a spin-echo EPI sequence. Key parameters of the acquisition were: TR = 4000 ms, TE = 72.7 ms, FOV = 220 * 220 mm, matrix = 110*110, slice thickness = 2mm, slice gap = 0, number of slice = 35. The acquired data was reconstructed using SENSE with the sensitivity map acquired from a separate low-resolution GRE sequence. Compared with the conventional SLR pulse, SLfRank has a reduction of 19\% energy with similar image quality.}

\section{Discussion}

Our results show that joint optimization of CK polynomials can produce pulses with reduced energy and more accurate phase profiles. In our opinion, reduced energy is the main feature of SLfRank pulses. Because SAR is proportional to pulse energy, SLfRank pulses can potentially accelerate acquisition time for SAR-limited sequences. In addition, there is a slight reduction of peak pulse amplitude, which can \revised{\rnum{4.6}be} useful to prevent overflow in power amplifiers for simultaneous multi-slice imaging. \revised{\rnum{1.2}Note that peak amplitude reduction is not guaranteed in all cases. In Supplementary Materials, we see that peak RF amplitude can increase for saturation pulses when designed with SLfRank.}

\revised{\rnum{1.3}
The main reason SLfRank has lower energy than SLR is that it jointly minimizes the energy as the objective for both CK polynomials. The optimal design  for a single polynomial in SLR is an equi-ripple filter (such as using Remez), which oscillates and touches the upper and lower constraint boundaries. SLfRank, on the other hand, does not have to be equi-ripple. And the results show that the minimum energy pulse profile usually touches the constraint boundaries near the edges, and then stays along one side without oscillating. This allows SLfRank to satisfy the profile constraints with lower energy. Note that SLfRank directly imposes time band-width and the desired magnetization profile constraints in the design. Therefore, the resulting pulse must satisfy the specified parameters. }

The accurate control over the transverse magnetization phase in SLfRank allows us to obtain a flatter phase response for slice excitation. However, this only provides marginal benefit in practice. As signals are contributed by summing across the excited slice, variation in the slice profile does not affect the resulting signal-to-noise ratio much. The SLfRank algorithm can still be useful for other applications, where pulse designers want to design more exotic phase profiles. \revised{\rnum{2.8}In particular, the current SLfRank algorithm can readily be used for quadratic phase and multiband designs, as we only need to specify desired magnetization profiles and appropriate constraints. Root flipping can potentially be a direction to explore, where we further reduce peak RF amplitude. In particular, we could apply root flipping on one of the CK polynomials after obtaining them from SLfRank. In general, we believe SLfRank could be a drop-in replacement for SLR in most applications.}

The proposed convex program can find a globally optimal solution for most examples in this manuscript. However, for the minimum phase excitation pulse design, there is still a slight gap between the solutions from the convex program and the rank-constrained problem. It is not clear to us whether this is a fundamental gap for such pulses. In particular, we observe that the convex program finds a rank-one solution when the ripple constraint is set to \(2\%\). It is possible that there is a regime where the convex program produces globally optimal pulses.

Compared to the original SLR algorithm, SLfRank takes much longer to compute. On a workstation with two 16-core Intel Xeon Silver 4216 processors, the original SLR algorithm takes less than a second to run, whereas the SLfRank algorithm takes around two minutes. Because most pulse designs are not done online on scanners, we believe the running time of SLfRank is reasonable. But there are also several directions to improve its computation time, including leveraging the fast Fourier transform in the iterative algorithm and using GPUs. \revised{\rnum{4.3}To accelerate the convergence of the iterative algorithm, we can potentially use SLR pulses to initialize the convex program. We can also explore a multi-level design where we use SLfRank CK polynomials discretized on a coarser resolution for initialization. For example, we can first solve the problem for $n=16$ and up-interpolate the result to $n=64$ as initialization.}

Finally, one limitation of both the SLR and SLfRank algorithms is that they assume the initial magnetization starts from equilibrium except for the special case of spin-echo refocusing with crusher gradients. \revised{In particular, the SLR transform assumes the initial CK parameters to be $\alpha_0 = 1$ and $\beta_0 = 0$. With arbitrary $\alpha_0$ and $\beta_0$, let us define the resulting CK parameters after the $n$th hard pulse to be $\tilde{\alpha}_n$ and $\tilde{\beta}_n$, then they are related to the original CK parameters with equilibrium as starting point as:
\begin{align*}
    \begin{pmatrix}
        \tilde{\alpha}_{n} \\
        \tilde{\beta}_{n}
    \end{pmatrix} 
    = 
    \begin{pmatrix}
    \alpha_n  &  -\beta_n^* \\
    \beta_n &  \alpha_n
  \end{pmatrix}
  \begin{pmatrix}
    1  & 0 \\
    0 &   z^{-1}
  \end{pmatrix} 
  \begin{pmatrix}
        \alpha_{0} \\
        \beta_{0}
    \end{pmatrix} .
\end{align*}
To generalize to arbitrary starting point, the transform or the design process should be changed accordingly to form a complete SLR algorithm. 
}
\section{Conclusion}

We have shown an improved SLR design process that can jointly solve for the CK polynomial pair. The new design can specify constraints directly on magnetization profiles, and optimize both CK polynomials to minimize pulse energy. The pulses in general have lower energy and fewer discontinuities. They also have more accurate phase responses when compared to the original SLR pulses. With lower energy pulses, the SLfRank algorithm can potentially accelerate SAR-limited sequences. Moreover, it allows users to design arbitrary excitation phase profiles, which opens up new research opportunities.

\bibliography{references}

\clearpage
\newpage

\section*{Supplementary Material}
\label{sec:supp_material}

\subsection*{RF Pulses with TBW=4 and n=64}
\label{ssec:tb4}

\revised{\rnum{1.2}Here we compare SLR and SLfRank pulses with time bandwidth (TBW) = 4 and $n=64$.}

\begin{figure}[h]
    \centering
    \includegraphics[width=\linewidth]{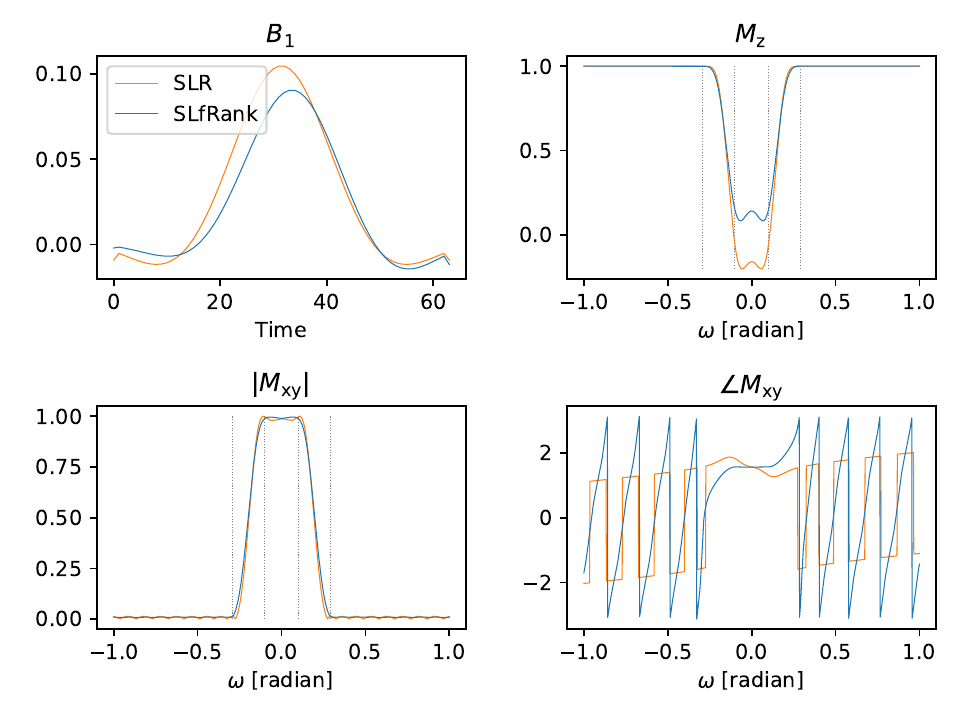}
    \caption{Linear phase excitation pulses and their magnetization profiles after gradient refocusing with TBW=4. Pulse energy is reduced from 0.157 to 0.114 (27.4\%) and peak is reduced from 0.104 to 0.090 (13.5\%).}
    \label{fig:tb4_excitation}
\end{figure}

\begin{figure}[h]
    \centering
    \includegraphics[width=\linewidth]{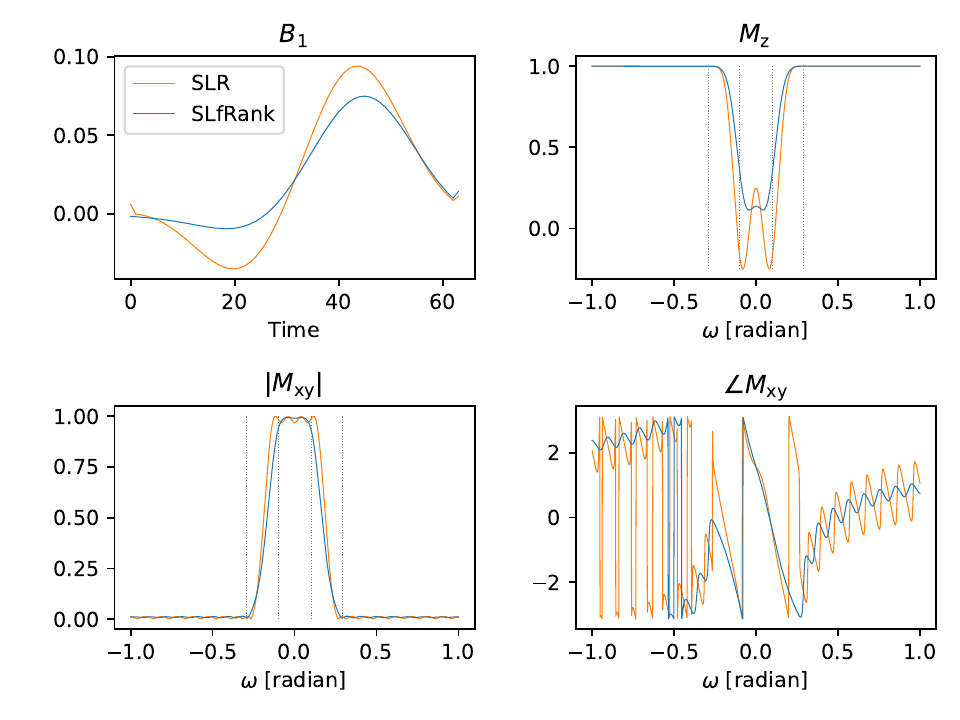}
    \caption{Minimum phase excitation pulses with TBW=4. Pulse energy is reduced from 0.143 to 0.089 (37.8\%) and peak is reduced from 0.094 to 0.075 (20.2\%).}
    \label{fig:tb4_ex_min}
\end{figure}

\begin{figure}[h]
    \centering
    \includegraphics[width=\linewidth]{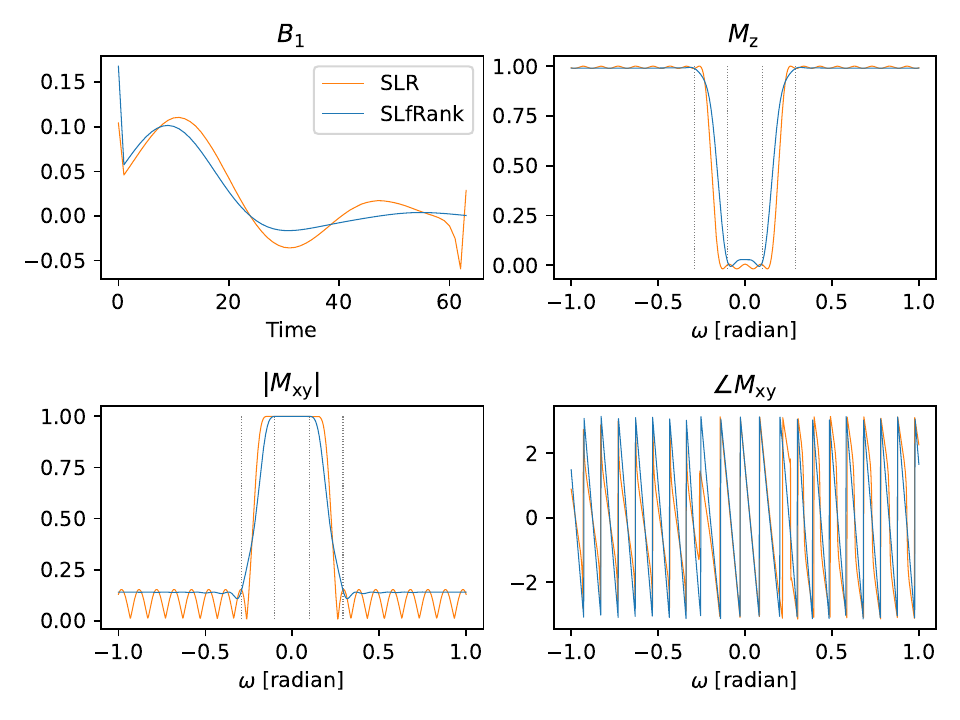}
    \caption{Maximum phase saturation pulses with TBW=4. Pulse energy is reduced from 0.179 to 0.157 (12.3\%). Peak is increased from 0.110 to 0.168 (52.7\%).}
    \label{fig:tb4_sat_max}
\end{figure}

\begin{figure}[h]
    \centering
    \includegraphics[width=\linewidth]{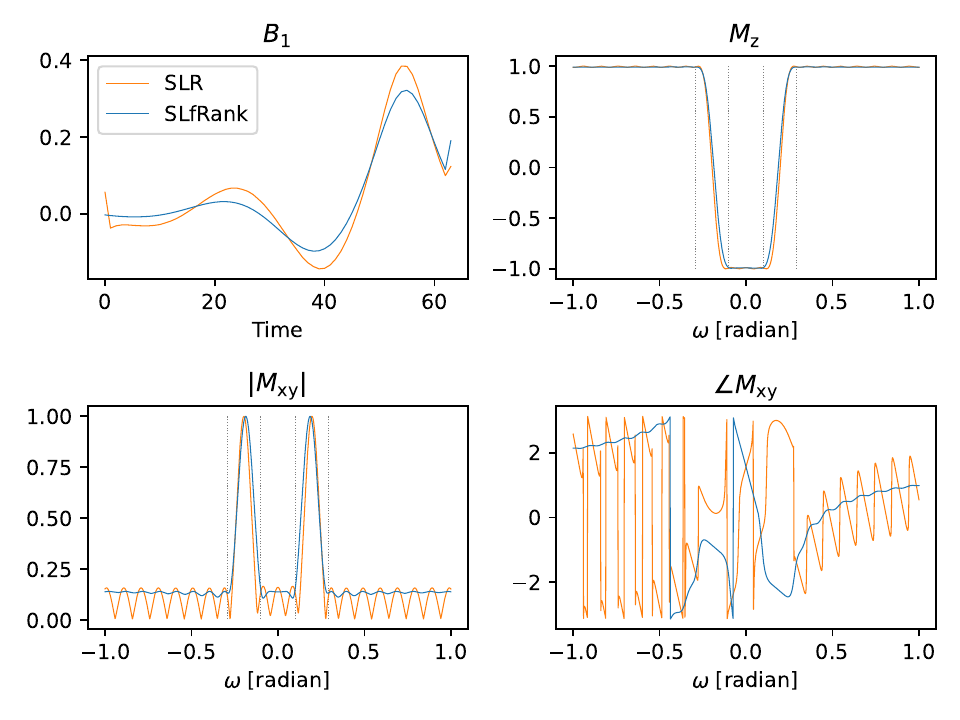}
    \caption{Minimum phase inversion pulses with TBW=4. Pulse energy is reduced from 1.338 to 0.992 (25.9\%) and peak is reduced from 0.385 to 0.322 (16.4\%).}
    \label{fig:tb4_inv_min}
\end{figure}

\begin{figure}[h]
    \centering
    \includegraphics[width=\linewidth]{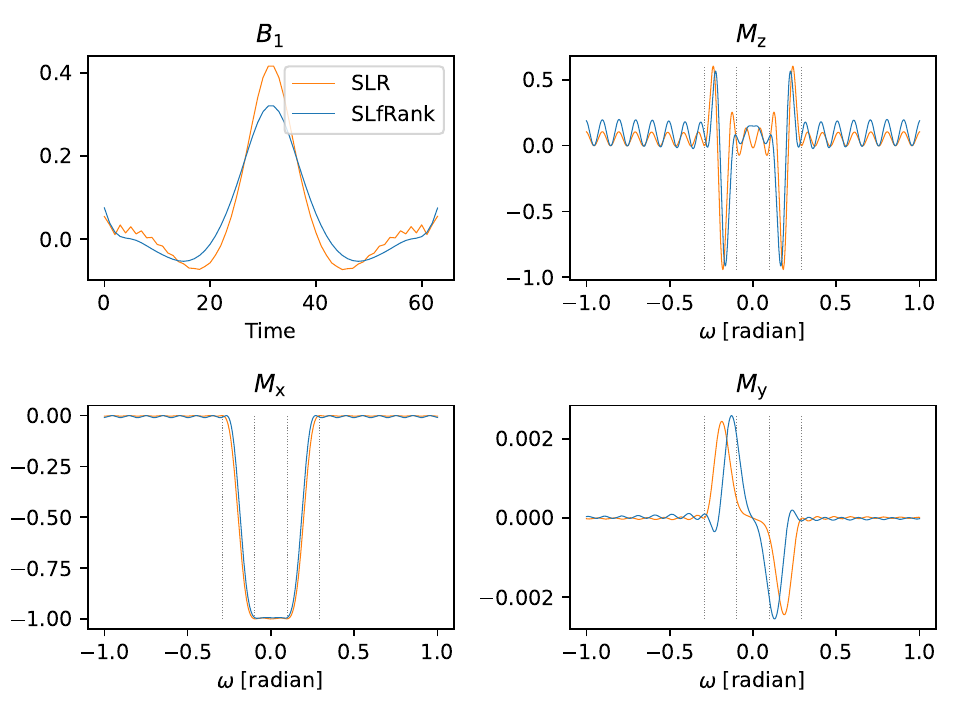}
    \caption{Zero phase spin-echo refocusing pulses with TBW=4. Pulse energy is reduced from 1.289 to 0.932 (27.7\%) and peak is reduced from 0.416 to 0.320 (23.1\%).}
    \label{fig:tb4_se_linear}
\end{figure}

\clearpage

\subsection*{RF Pulses with TBW=10 and n=64}
\label{ssec:tb10}

\revised{\rnum{1.2}Here we compare SLR and SLfRank pulses with TBW = 10 and $n=64$.}

\begin{figure}[h]
    \centering
    \includegraphics[width=\linewidth]{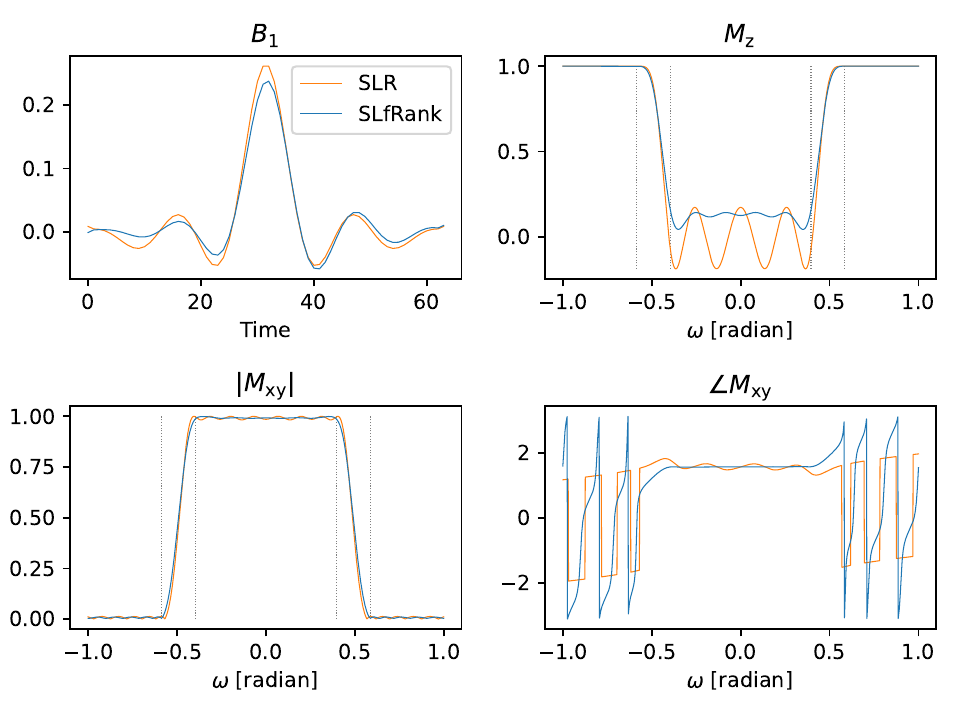}
    \caption{Linear phase excitation pulses and their magnetization profiles after gradient refocusing with TBW=10. Pulse energy is reduced from 0.259 to 0.212 (18.1\%) and peak is reduced from 0.169 to 0.153 (9.47\%).}
\end{figure}

\begin{figure}[h]
    \centering
    \includegraphics[width=\linewidth]{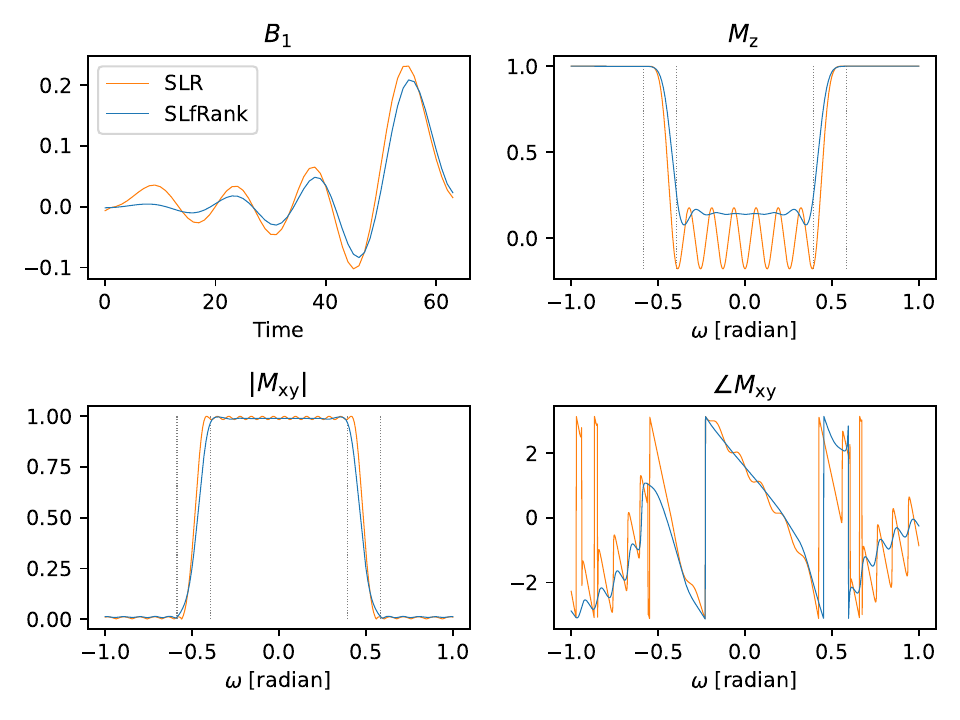}
    \caption{Minimum phase excitation pulses with TBW=10. Pulse energy is reduced from 0.259 to 0.194 (25.1\%) and peak is reduced from 0.149 to 0.132 (17.0\%).}
\end{figure}

\begin{figure}[h]
    \centering
    \includegraphics[width=\linewidth]{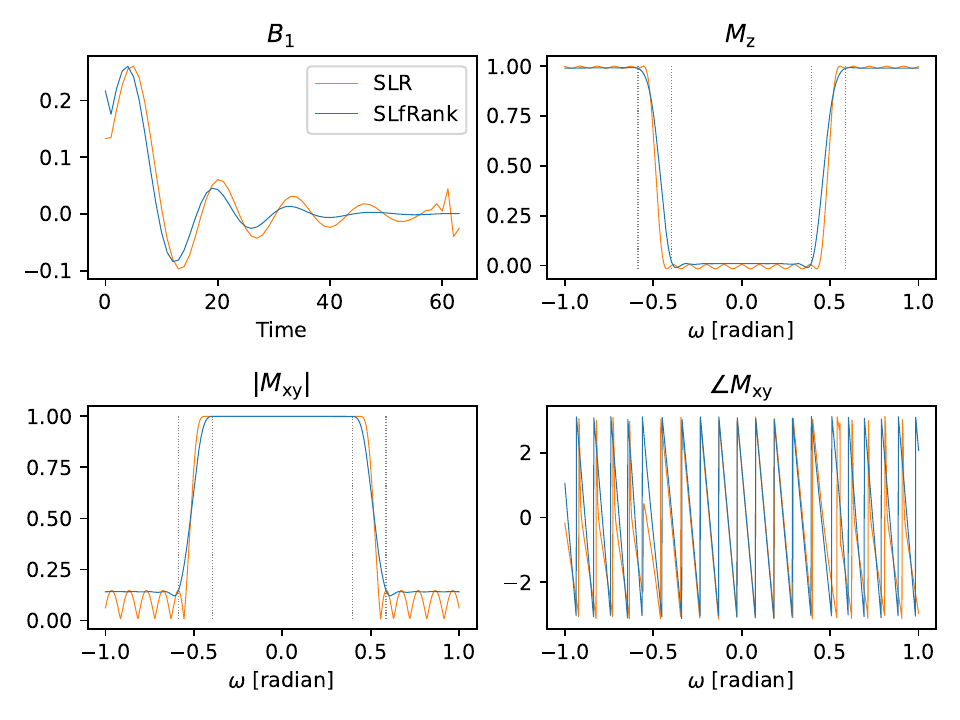}
    \caption{Maximum phase saturation pulses with TBW=10. Pulse energy is reduced from 0.295 to 0.275 (6.78\%). Peak is increased from 0.167 to 0.186 (12.7\%).}
\end{figure}

\begin{figure}[h]
    \centering
    \includegraphics[width=\linewidth]{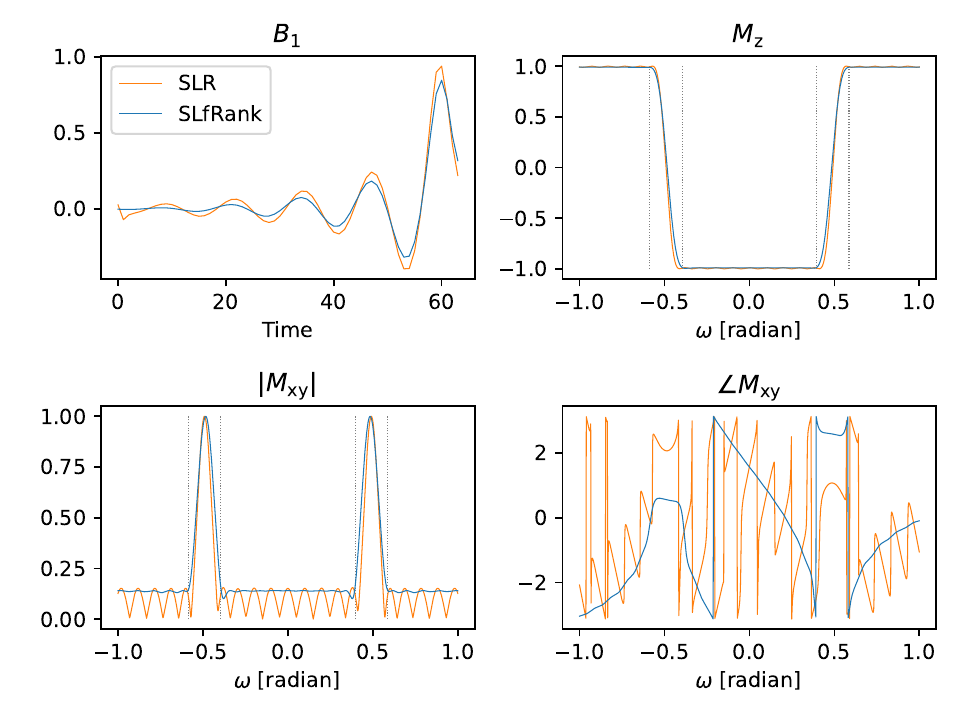}
    \caption{Minimum phase inversion pulses with TBW=10. Pulse energy is reduced from 2.440 to 1.910 (21.7\%) and peak is reduced from 0.624 to 0.549 (12.0\%).}
\end{figure}

\begin{figure}[h]
    \centering
    \includegraphics[width=\linewidth]{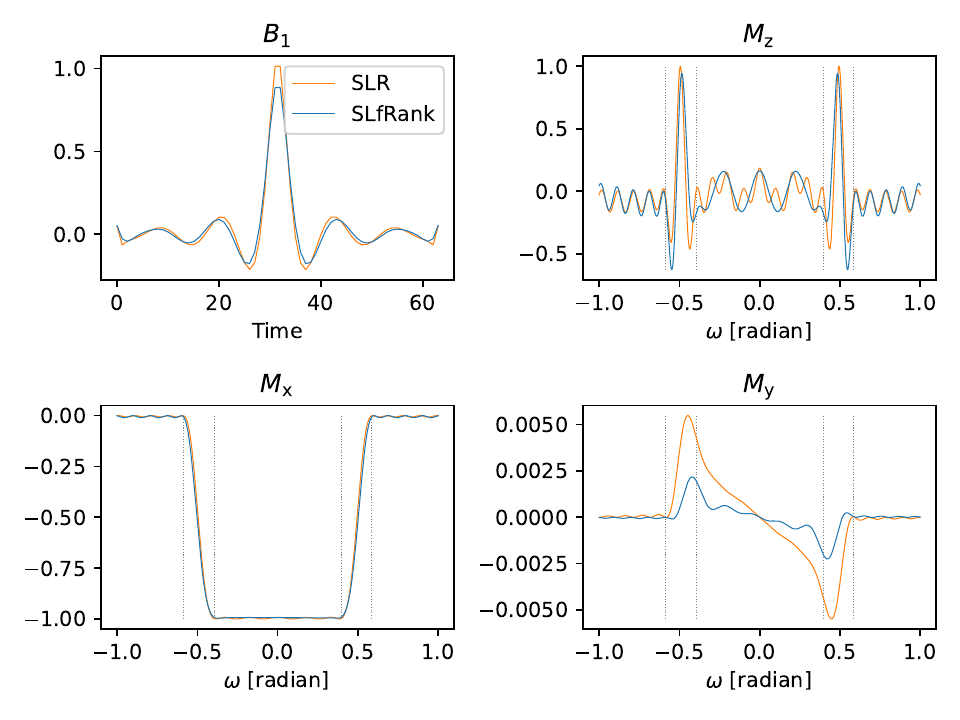}
    \caption{Zero phase spin-echo refocusing pulses with TBW=10. Pulse energy is reduced from 2.263 to 1.794 (20.7\%) and peak is reduced from 0.687 to 0.579 (15.7\%).}
\end{figure}

\clearpage

\subsection*{RF Pulses with TBW=8 and n=100}
\label{ssec:n100}

\revised{\rnum{1.2}Here we compare SLR and SLfRank pulses with TBW = 8 and $n=100$.}

\begin{figure}[h]
    \centering
    \includegraphics[width=\linewidth]{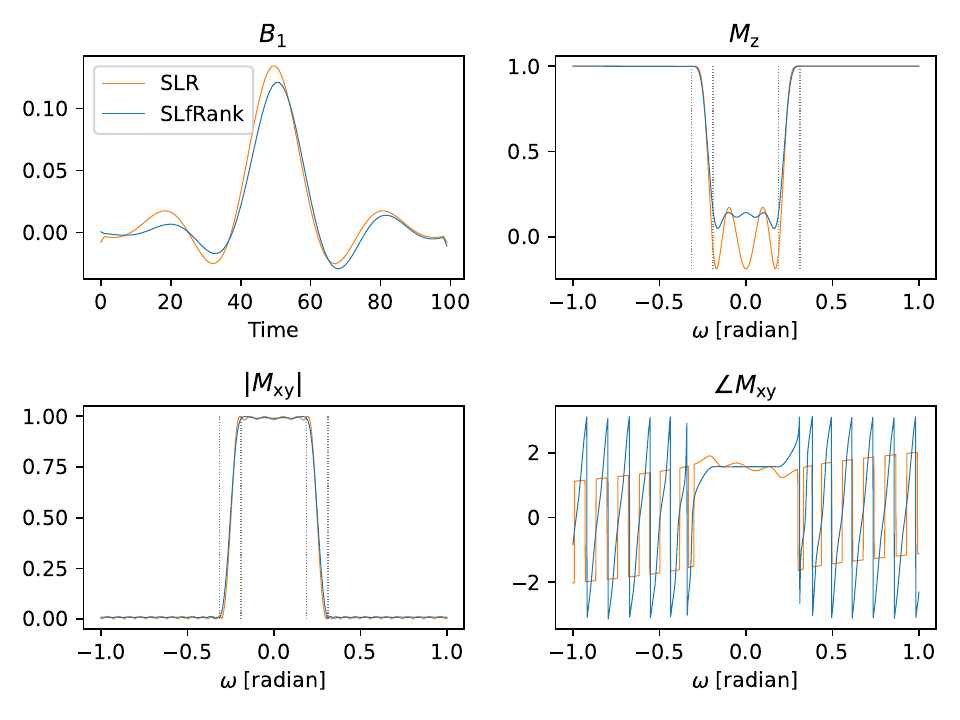}
    \caption{Linear phase excitation pulses and their magnetization profiles after gradient refocusing with $n=100$. Pulse energy is reduced from 0.204 to 0.166 (18.6\%) and peak is reduced from 0.134 to 0.121 (9.70\%).}
    \label{fig:n100_ex_linear}
\end{figure}

\begin{figure}[h]
    \centering
    \includegraphics[width=\linewidth]{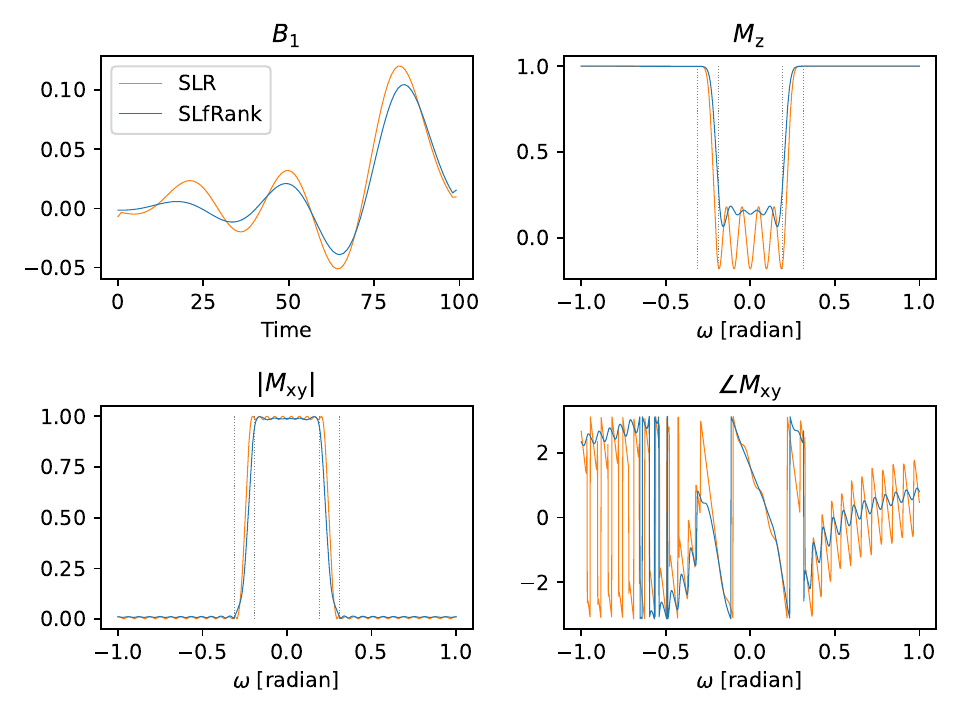}
    \caption{Minimum phase excitation pulses with $n=100$. Pulse energy is reduced from 0.203 to 0.148 (27.1\%) and peak is reduced from 0.120 to 0.105 (12.5\%).}
    \label{fig:n100_ex_min}
\end{figure}

\begin{figure}[h]
    \centering
    \includegraphics[width=\linewidth]{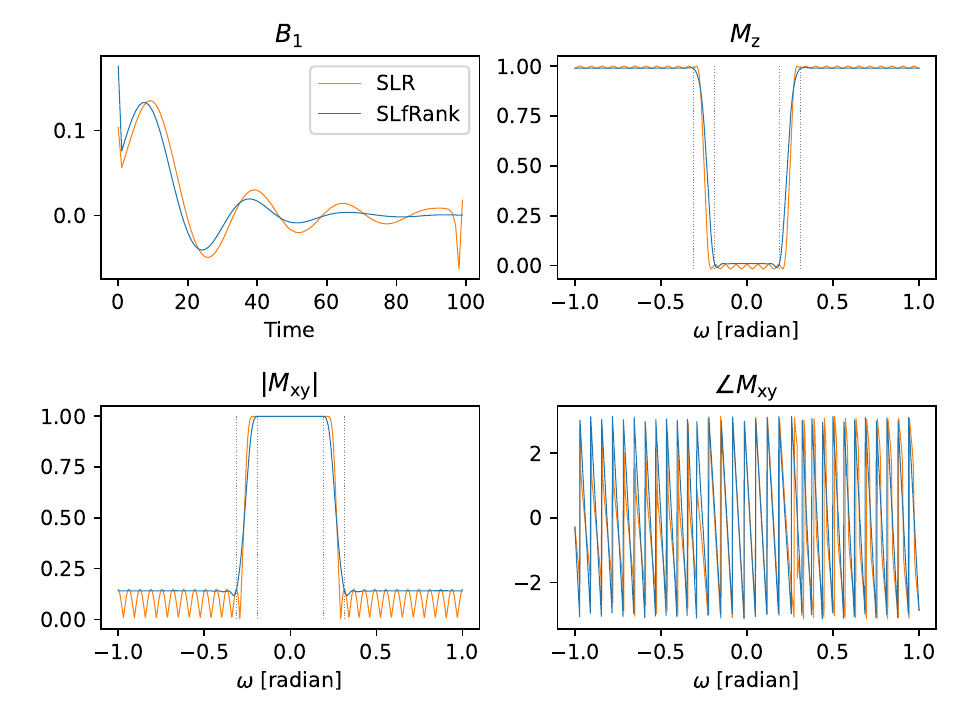}
    \caption{Maximum phase saturation pulses with $n=100$. Pulse energy is reduced from 0.229 to 0.220 (3.94\%). Peak is increased from 0.135 to 0.176 (30.4\%).}
    \label{fig:n100_sat_max}
\end{figure}

\begin{figure}[h]
    \centering
    \includegraphics[width=\linewidth]{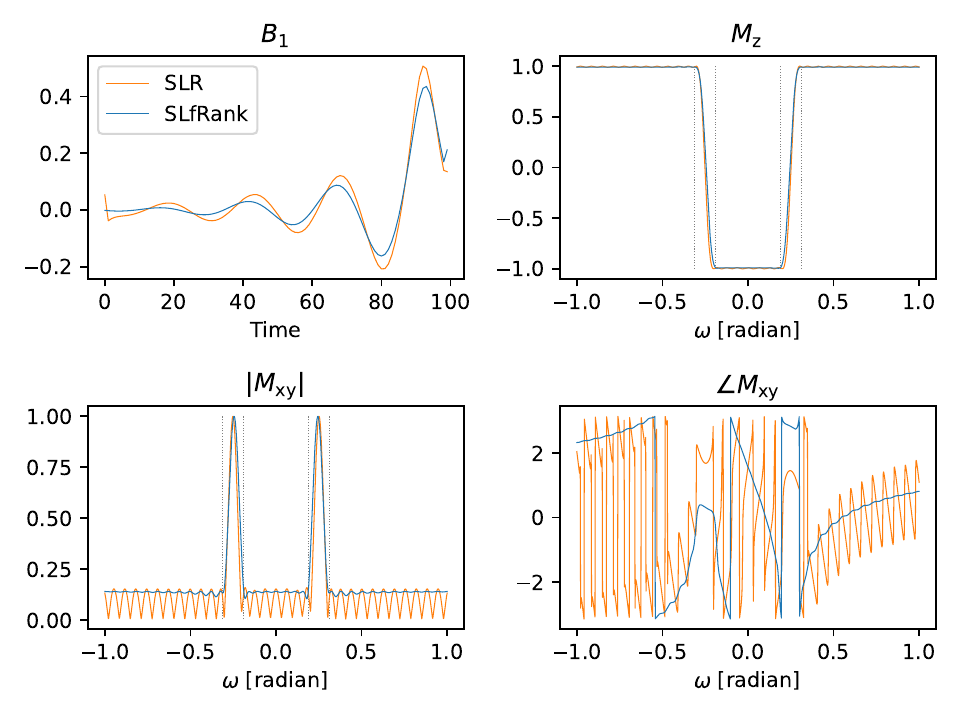}
    \caption{Minimum phase inversion pulses with $n=100$. Pulse energy is reduced from 1.938 to 1.489 (23.2\%) and peak is reduced from 0.507 to 0.435 (14.2\%).}
    \label{fig:n100_inv_min}
\end{figure}

\begin{figure}[h]
    \centering
    \includegraphics[width=\linewidth]{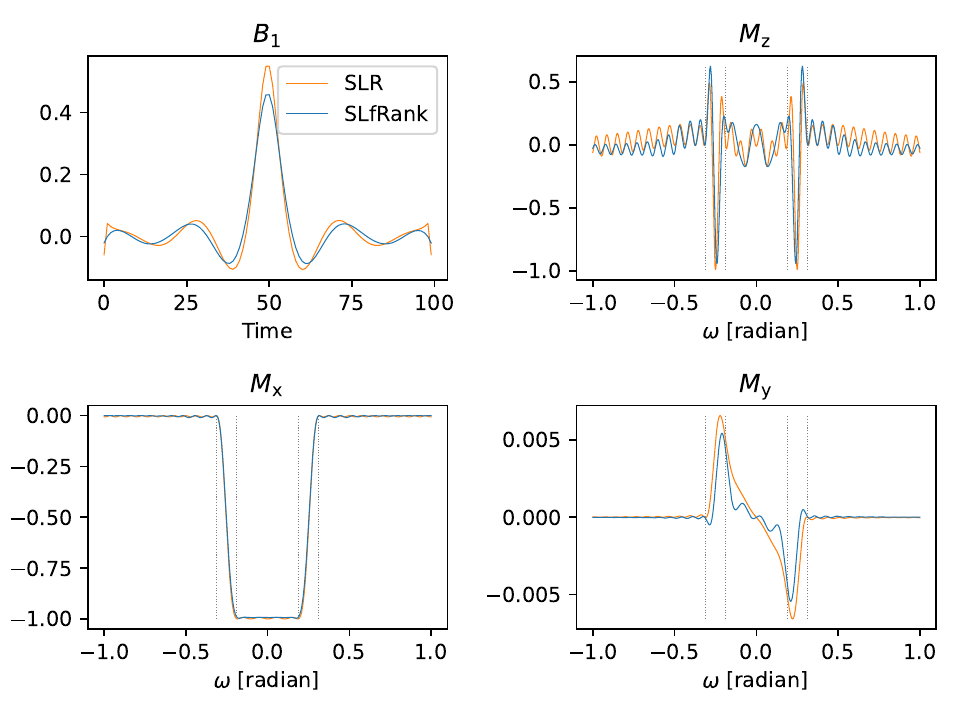}
    \caption{Zero phase spin-echo refocusing pulses with $n=100$. Pulse energy is reduced from 1.778 to 1.390 (21.8\%) and peak is reduced from 0.550 to 0.458 (16.7\%).}
    \label{fig:n100_se_linear}
\end{figure}

\clearpage
\subsection*{Proof Sketch for Equation \eqref{eq:enery_con}}

Here we provide a proof sketch to show that for any matrix $\mathbf{M} \in \mathbb{C}^{n \times n}$, the following two equations are equivalent:
\[
   1 = \bm{\psi}^*(e^{\imath \omega})  \mathbf{M} \bm{\psi}(e^{\imath \omega}) \text{ for all } \omega
\]
where \(\bm{\psi}(z) = 
\begin{pmatrix}
1 &
 e^{\imath \omega} &
\ldots &
 e^{\imath \omega n}
 \end{pmatrix}^T
 \),
 and
\[
\sum_{i, j : i - j = k} (\mathbf{M})_{ij} = \begin{cases}
    1,& \text{if } k = 0\\
    0,              & \text{otherwise}
\end{cases},
\]
for \(k = -n, -n + 1, \ldots, n\).

Let us define \(\mathbf{1}\) to be an all-one vector and  \( \odot \) be the element-wise product, then
\[
    \bm{\psi}^*(z)  \mathbf{M} \bm{\psi}(e^{\imath \omega}) = \mathbf{1}^*  \left( \mathbf{M} \odot \left[ \bm{\psi}(e^{\imath \omega}) \bm{\psi}^*(e^{\imath \omega}) \right]\right)  \mathbf{1}.
\]

A special structure of the matrix $\bm{\psi}(e^{\imath \omega}) \bm{\psi}^*(e^{\imath \omega})$ is that each of its sub-diagonal has the same value of a phase exponential. That is,
\[
\left( \bm{\psi}(e^{\imath \omega}) \bm{\psi}^*(e^{\imath \omega}) \right)_{ij} = e^{\imath \omega (i - j)}
\]
for \(i, j = 1, \ldots, n\).

Hence $\mathbf{1}^*  \mathbf{M} \odot \left[ \bm{\psi}(e^{\imath \omega}) \bm{\psi}^*(e^{\imath \omega}) \right] \mathbf{1}$ is the same as the inner product between a vector \(\mathbf{v}\) formed by diagonally summing \(\mathbf{M}\) and a linear phase vector \(\bm{\phi}(e^{\imath \omega}) = [e^{-\imath \omega n}, \ldots, e^{\imath \omega (n - 1)}, e^{\imath \omega n}]^T \). In particular, \(\mathbf{v}\) is defined as
\[
(\mathbf{v})_{k + n + 1} = \sum_{i, j : i - j = k} (\mathbf{M})_{ij}
\]
for \(k = -n, -n + 1, \ldots, n\).

Finally, to satisfy \( 1 = \bm{\psi}^*(e^{\imath \omega})  \mathbf{M} \bm{\psi}(e^{\imath \omega})\) for all \(\omega\), one must have \((\mathbf{v})_{n + 1}\) being one and all other values being zero. This is equivalent to have main diagonal sum of \(\mathbf{M}\) being one, but all other sub-diagonal sum being zero.

\clearpage
\subsection*{Convergence Plot}
To show the convergence process of the algorithm, the error rank of different RF pulse design types was plotted, which was defined as: 
\[
    error\ rank = \left\| \mathbf{P} -
      \begin{pmatrix}
          \mathbf{a} \\
          \mathbf{b}
      \end{pmatrix}
      \begin{array}{@{}c@{}}
          \begin{pmatrix}
              \mathbf{a}^* & \mathbf{b}^*
          \end{pmatrix} \\
        \mathstrut 
      \end{array}
      \right\|_2
\]
The plot shows that the convex program always converges, however, minimum phase designs take more iterations. 
\begin{figure}[h]
    \centering
    \includegraphics[width=\linewidth]{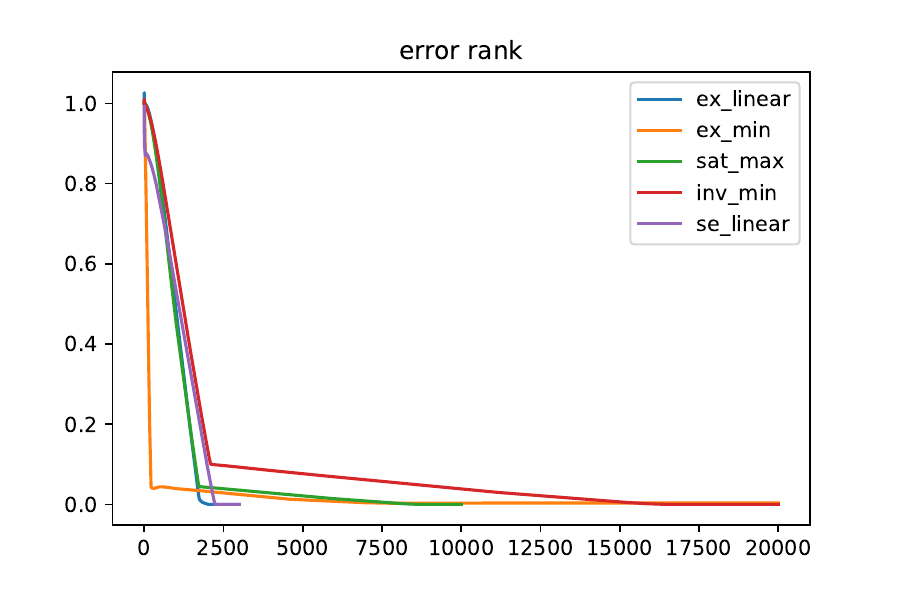}
    \caption{Convergence plot of the different RF pulse designs using the proposed algorithm. In most cases, the algorithm will converge around 2000 iterations.}
    \label{fig:error_rank_plot}
\end{figure}

\end{document}